\documentclass[twocolumn]{aastex63}
\usepackage{natbib}

\shorttitle{G208-N2}
\shortauthors{Hirano et al.}
\graphicspath{{./}{figures/}}

\begin{document}

\title{ALMA Survey of Orion Planck Galactic Cold Clumps (ALMASOP): 
Discovery of an extremely dense and compact object embedded in the prestellar core G208.68-19.92-N2}

\correspondingauthor{Naomi Hirano}
\email{hirano@asiaa.sinica.edu.tw}

\author[0000-0001-9304-7884]{Naomi Hirano}
\affiliation{Academia Sinica Institute of Astronomy and Astrophysics,
11F of Astronomy-Mathematics Building, AS/NTU,
No.1, Sec. 4, Roosevelt Rd, Taipei 10617, Taiwan, R.O.C.}

\author[0000-0002-4393-3463]{Dipen Sahu}
\affiliation{Physical Research laboratory, Navrangpura, Ahmedabad, Gujarat 380009, India}
\affiliation{Academia Sinica Institute of Astronomy and Astrophysics,
11F of Astronomy-Mathematics Building, AS/NTU,
No.1, Sec. 4, Roosevelt Rd, Taipei 10617, Taiwan, R.O.C.}

\author[0000-0012-3245-1234]{Sheng-Yaun Liu}
\affiliation{Academia Sinica Institute of Astronomy and Astrophysics,
11F of Astronomy-Mathematics Building, AS/NTU,
No.1, Sec. 4, Roosevelt Rd, Taipei 10617, Taiwan, R.O.C.}

\author[0000-0002-5286-2564]{Tie Liu}
\affiliation{Shanghai Astronomical Observatory, Chinese Academy of Sciences, 80 Nandan Road, Shanghai 200030, People's Republic of China}

\author[0000-0002-8149-8546]{Ken'ichi Tatematsu}
\affiliation{Nobeyama Radio Observatory, National Astronomical Observatory of Japan,
National Institutes of Natural Sciences,
462-2 Nobeyama, Minamimaki, Minamisaku, Nagano 384-1305, Japan}

\author[0000-0002-2338-4583]{Somnath Dutta}
\affiliation{Academia Sinica Institute of Astronomy and Astrophysics,
11F of Astronomy-Mathematics Building, AS/NTU,
No.1, Sec. 4, Roosevelt Rd, Taipei 10617, Taiwan, R.O.C.}

\author[0000-0003-1275-5251]{Shanghuo Li}
\affiliation{Max Planck Institute for Astronomy, K\"onigstuhl 17, 69117 Heidelberg, Germany}

\author[0000-0002-3024-5864]{Chin-Fei Lee}
\affiliation{Academia Sinica Institute of Astronomy and Astrophysics,
11F of Astronomy-Mathematics Building, AS/NTU,
No.1, Sec. 4, Roosevelt Rd, Taipei 10617, Taiwan, R.O.C.}

\author[0000-0001-8077-7095]{Pak Shing Li}
\affiliation{Shanghai Astronomical Observatory, Chinese Academy of Sciences, 80 Nandan Road, Shanghai 200030, People's Republic of China}

\author[0000-0002-1369-1563]{Shih-Ying Hsu}
\affiliation{National Taiwan University (NTU), Taiwan, R.O.C.}
\affiliation{Academia Sinica Institute of Astronomy and Astrophysics,
11F of Astronomy-Mathematics Building, AS/NTU,
No.1, Sec. 4, Roosevelt Rd, Taipei 10617, Taiwan, R.O.C.}

\author[0000-0002-6868-4483]{Sheng-Jun Lin}
\affiliation{Academia Sinica Institute of Astronomy and Astrophysics,
11F of Astronomy-Mathematics Building, AS/NTU,
No.1, Sec. 4, Roosevelt Rd, Taipei 10617, Taiwan, R.O.C.}

\author[0000-0002-6773-459X]{Doug Johnstone}
\affiliation{NRC Herzberg Astronomy and Astrophysics, 5071 West Saanich Rd, Victoria, BC, V9E 2E7, Canada}
\affiliation{Department of Physics and Astronomy, University of Victoria, Victoria, BC, V8P 5C2, Canada}

\author[0000-0002-9574-8454]{Leonardo Bronfman}
\affiliation{Departamento de Astronomía, Universidad de Chile, Camino el Observatorio 1515, Las Condes, Santiago, Chile}

\author[0000-0002-9774-1846]{Huei-Ru Vivien Chen}
\affiliation{Institute of Astronomy and Department of Physics, National Tsing Hua University, Hsinchu, 30013, Taiwan, R.O.C.}

\author[0000-0002-5881-3229]{David J. Eden}
\affiliation{Armagh Observatory and Planetarium, College Hill, Armagh, BT61 9DB, UK.}

\author[0000-0002-4336-0730] {Yi-Jehng Kuan}
\affiliation{Department of Earth Sciences, National Taiwan Normal University, Taipei, Taiwan (R.O.C.)}
\affiliation{Academia Sinica Institute of Astronomy and Astrophysics, 11F of Astronomy-Mathematics Building, AS/NTU, No.1, Sec. 4, Roosevelt Rd, Taipei 10617, Taiwan, R.O.C.}

\author[0000-0003-4022-4132]{Woojin Kwon}
\affiliation{Department of Earth Science Education, Seoul National University, 1 Gwanak-ro, Gwanak-gu, Seoul 08826, Republic of Korea}
\affiliation {SNU Astronomy Research Center, Seoul National University, 1 Gwanak-ro, Gwanak-gu, Seoul 08826, Republic of Korea}

\author[0000-0002-3179-6334]{Chang Won Lee}
\affiliation{Korea Astronomy and Space Science Institute, 776 Daedeokdae-ro, Yuseong-gu, Daejeon 34055, Republic of Korea }
\affiliation{University of Science and Technology, Korea (UST), 217 Gajeong-ro, Yuseong-gu, Daejeon 34113, Republic of Korea}

\author[0000-0003-3343-9645]{Hong-Li Liu}
\affiliation{Department of Astronomy, Yunnan University, and Key Laboratory of Particle Astrophysics of Yunnan Province, Kunming, 650091, People’s Republic of China}

\author[0000-0002-6529-202X]{Mark G. Rawlings}
\affiliation{Gemini Observatory/NSF’s NOIRLab, 670 N. A’ohoku Place, Hilo, Hawai’i, 96720, USA}

\author[0000-0002-1469-6323]{Isabelle Ristorcelli}
 \affiliation{ Universit\'e de Toulouse, UPS-OMP, IRAP, F-31028 Toulouse cedex 4, France}

 \author[0000-0003-1665-6402]{Alessio Traficante}
 \affiliation{IAPS-INAF, via Fosso del Cavaliere 100, I-00133, Rome, Italy}







\begin{abstract}

The internal structure of the prestellar core G208.68-19.02-N2 (G208-N2) in the Orion Molecular Cloud 3 (OMC-3) region has been studied with the Atacama Large Millimeter/submillimeter Array (ALMA).
The dust continuum emission revealed a  filamentary structure with a length of $\sim$5000 au and an average H$_2$ volume density of $\sim$6 $\times$ 10$^7$ cm$^{-3}$. 
At the tip of this filamentary structure, there is a compact object, which we call a ``nucleus", with a radius of $\sim$150--200 au and a mass of $\sim$0.1 M$_{\odot}$.
 The nucleus has a central density of $\sim$2 $\times$ 10$^9$ cm$^{-3}$ with a radial density profile of  $r^{-1.87{\pm}0.11}$.
The density scaling of the nucleus is $\sim$3.7 times higher than that of the singular isothermal sphere. 
 This as well as the very low virial parameter of 0.39 suggest that the gravity is dominant over the pressure everywhere in the nucleus.
However, there is no sign of CO outflow localized to this nucleus. 
The filamentary structure is traced by the N$_2$D$^+$ 3--2 emission, but not by the C$^{18}$O 2--1 emission, implying the significant CO depletion due to high density and cold temperature.
Toward the nucleus, the N$_2$D$^+$ also shows the signature of depletion.
This could imply either the depletion of the parent molecule, N$_2$, or the presence of the embedded very-low luminosity central source that could sublimate the CO in the very small area.
The nucleus in G208-N2 is considered to be a prestellar core on the verge of first hydrostatic core (FHSC) formation or a candidate for the FHSC.
\end{abstract}

\keywords{Molecular clouds (1072); Collapsing clouds (267); Star forming regions (1565); Star formation (1569); Astrochemistry (75); Early stellar evolution (434)}


\section{Introduction} \label{sec:intro}

Dense, cold, and gravitationally unstable  starless cores prior to the onset of star formation, so-called ``prestellar cores'' are ideal targets for studying the initial condition of protostellar collapse \citep[e.g.][]{diF07,War07}.
Studying the physical conditions such as temperature, density, and gas kinematics of prestellar cores is essential to understand how the cores collapse and initiate star formation.

The Planck Galactic Cold Clumps (PGCCs: \cite{Pla16}) are considered as the ideal targets for studying the initial condition of star formation due to their low dust temperatures of 6--20 K.
Using the 15m James Clerk Maxwell Telescope (JCMT) equipped with the Submillimeter Common User Bolometer Array-2 (SCUBA-2), the legacy survey  ``SCOPE: SCUBA-2 Continuum Observations of Pre-protostellar Evolution'' has observed 1235 PGCCs at an angular resolution of 14\farcs4, and catalogued $\sim$3500 dense cores \citep{Liu18,Ede19}. 
In the Orion molecular cloud complex including Orion A, Orion B, and $\lambda$ Orionis regions, the SCOPE survey identified 119 dense cores \citep{Yi18}.
These cores are further studied in molecular lines at the 3 mm band using the Nobeyama Radio Observatory (NRO) 45m telescope \citep{Kim20,Tat17,Tat21,Tat22}.
Using the N$_2$D$^+$ 1--0 intensity observed with the NRO 45 m telescope, 72 Orion dense cores have been selected; 23 out of 72 have no mid-infrared source in the {\it Wide-Field Infrared Survey Explorer} (WISE) catalog \citep{Wri10}, while 49 have WISE sources.
The selected starless cores show intense N$_2$D$^+$ emission, and are considered to be in the evolved starless stage close to the onset of star formation.
The sample of protostellar cores include the ones with and without intense N$_2$D$^+$ emission.
The former with intense N$_2$D$^+$ are considered to be in an earlier evolutionary stage than the latter without N$_2$D$^+$ emission.
Therefore, the selected 72 dense cores cover the evolutionary stages before and after the onset of star formation.
The ``ALMA Survey of Orion PGCCs (ALMASOP)'' project has observed these 72 dense cores at high angular resolution in the continuum and molecular lines at 1.3 mm in order to study the fragmentation and substructures of dense cores at the onset of star formation.
The description and the full sample of ALMASOP are presented in \citet{Dut20}.

\begin{figure}
\epsscale{1.2}
\plotone{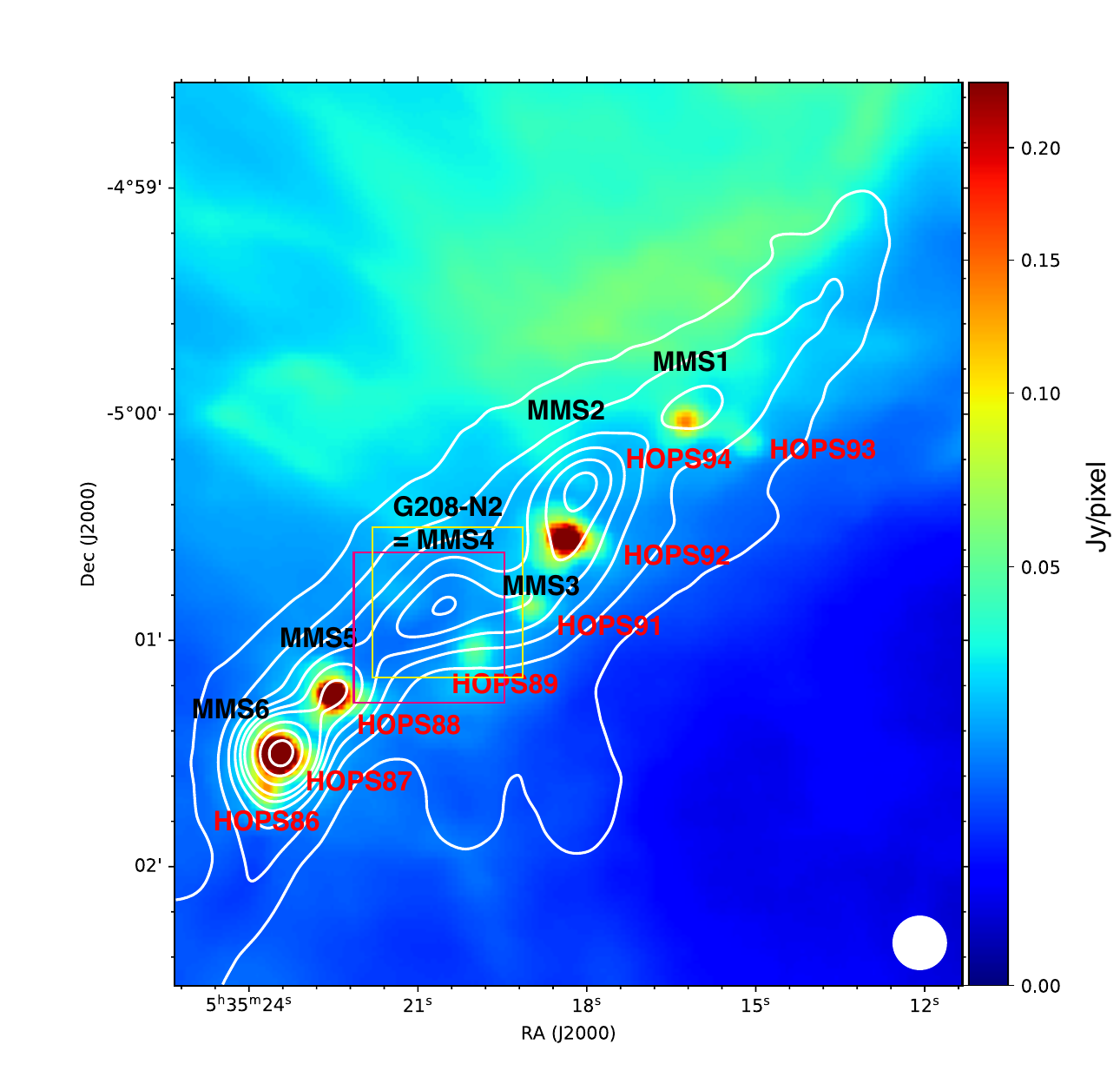}
\caption{850 $\mu$m continuum emission of the northern part of OMC-3 region including G208.68-19.20-north observed with the JCMT SCUBA-2 \citep{Yi18} in contours overlaid on the {\it Hershel} PACS 70 $\mu$m image in color.
Contours are drawn every 5 $\sigma$ step with the lowest contour level of 5 $\sigma$.
1 $\sigma$ level is 64.2 mJy beam$^{-1}$.
White circle in the bottom right denotes the JCMT beam (14\farcs0).
The names of the cores follow the nomenclature of \citet{Chi97} and the protostellar sources in the HOPS catalog are given in the black and red letters, respectively.
Magenta and yellow boxes denote the regions of Figure \ref{fig:cont}a and  \ref{fig:cont}c, respectively.
 \label{fig:PACS_SCUBA}}
\end{figure}

G208.68-19.20-N2 (G208-N2) is one of the SCUBA\textcolor{blue}{-2} cores identified in the PGCC G208.68-19.20 North \citep{Yi18}, which corresponds to the northern part of the Orion Molecular Cloud 3 (OMC-3).
The SCUBA-2 850 $\mu$m continuum image of the OMC-3 region including G208.68-19.20 North on top of the {\it Herschel} PACS 70 $\mu$m image from archive is shown in Figure \ref{fig:PACS_SCUBA}.
The distance to OMC-3 is estimated to be $\sim$390 pc \citep{Kou17,Kou18,Tob20}.
This core was also identified in the previous observations at 1.3 mm \citep[MMS4;][]{Chi97}, 350 $\mu$m \citep[CSO9; ][]{Lis98}, and 850 $\mu$m \citep[OriAN-535207-50053;][]{Nut07}, \citep[SMM5;][]{Tak13}.
The 850 $\mu$m image at 4\farcs5 resolution obtained with the Submillimeter Array (SMA) shows that this core is elongated along the northwest--southeast direction and surrounded by an envelope with a triangle shape \citep{Tak13}.
Figure \ref{fig:PACS_SCUBA} shows that there is an infrared source HOPS 89, which is classified as a flat spectrum source \citep{Meg12,Fur16}, to the southwest of the JCMT continuum peak of G208-N2 \citep{Yi18}.
However, this source is offset by $\sim$12\arcsec\ from the JCMT continuum peak, and is located outside of the triangle-shaped envelope observed at 850 $\mu$m \citep{Tak13} and 1.3 mm \citep{Dut20}.
Therefore, G208-N2 is likely to be starless.

The 3 mm line observations with the NRO 45m telescope have revealed that the N$_2$H$^+$ 1--0 and N$_2$D$^+$ 1--0 intensity values of this core are the brightest among the 119 cores in the Orion region (Kim et al. 2020).
The N$_2$D$^+$/N$_2$H$^+$ column density ratio in G208-N2 was estimated to be 0.11$\pm$0.01 \citep{Kim20}
, indicating that this core is in the late evolutionary stage of the starless phase where $N$(N$_2$D$^+$)/$N$(N$_2$H$^+$) $>$ 0.1 \citep{Cra07}, likely to be close to the onset of star formation.
This scenario is also supported by the 1.3 mm continuum observations of ALMASOP.
\citet{Sah21} has reported the detection of a compact and dense ($\sim$10$^7$ cm$^{-3}$) substructure in five starless cores including G208-N2 at 1\farcs2 resolution.
They found that these five cores are gravitationally unstable prestellar cores.
The 1.3 mm continuum emission from G208-N2 is the brightest among these five cores with substructures.

In this work, we utilize ALMASOP and archival ALMA data, and study the internal structure of G208-N2 using the continuum emission at 1.3 mm and 1.1 mm, and the molecular lines at 1.3 mm.
Thanks to the high resolution and high sensitivity of ALMA, the continuum images have revealed an embedded compact object having a radius of $\sim$200 au and a central density of ${\gtrsim}$10$^9$ cm$^{-3}$.
We discuss the physical properties and the evolutionary stage of this compact object.
We also study the chemical stratification in G208-N2 using the molecular line data of C$^{18}$O, N$_2$D$^+$, DCO$^+$, and H$_2$CO.

\section{Observations} \label{sec:obs}

\subsection{ALMASOP} \label{subsec:almasop}

The ALMA band 6 (1.3 mm) observations of G208-N2 were carried out as a part of the ALMASOP project in Cycle-6 (\#2018.1.00302.S; PI; Tie Liu).
The observations were executed in three different array configurations; 12m array C43-5 (TM1), 12m array C43-2 (TM2), and 7m array of the Atacama Compact Array (ACA).
The TM1, TM2, and ACA observations were conducted in three, one, and five execution blocks (EBs), respectively.
The projected baseline lengths of these three configurations are from 9 to 1090 k$\lambda$ in TM1, 9 to 240 k$\lambda$ in TM2, and 5 to 36 k$\lambda$ in ACA. 
Four spectral windows (SPWs) centered at the rest frequencies of 231.0 GHz, 233.0 GHz, 218.918 GHz, and 216.6178 GHz were used simultaneously.
Each SPW was set to a total bandwidth of 1.875 GHz and a spectral resolution of 1.129 MHz.
The corresponding velocity resolution was $\sim$1.5 km s $^{-1}$.
This spectral setting accommodates the $J=$2--1 of CO and C$^{18}$O, $J=$3--2 of N$_2$D$^+$, DCO$^+$, and DCN, three transitions of H$_2$CO, 3$_{0,3}$--2$_{0,2}$, 3$_{2,2}$--2$_{2,1}$, and 3$_{2,1}$--2$_{2,0}$, and the SiO $J=$5--4 lines.
The total bandwidth for the continuum was 7.5 GHz.
The details of the observational parameters are summarized in \citet{Dut20}.
The coordinates of the phase center for G208-N2 were $\alpha_{\rm ICRS}$ = 5$^{\rm h}$35$^{\rm m}$20.469$^{\rm s}$, $\delta_{\rm ICRS}$ = $-$5\arcdeg 00\arcmin  50\farcs394.

The visibility data of each EB were calibrated separately using the standard pipeline of CASA 5.4 \citep{McM07}.
The continuum visibility data were obtained by averaging the line-free channels of the four spectral windows.
The continuum emission was subtracted from the line data in the visibility domain.
The calibrated visibility data of TM1, TM2, and ACA were combined, and imaged using the TCLEAN task of CASA 5.5.
The maximum recoverable scale (MRS) was 38\farcs6.
In order to improve the signal-to-noise ratio, the molecular lines other than CO were imaged using the uvtaper of 0\farcs8 and the Briggs weighting with a robust of +0.5, which provided the synthesized beam size of 1\farcs1$\times$1\farcs0.
The 1.3 mm continuum and CO 2--1 line were imaged with and without uvtaper using the Briggs weighting with a robust of +0.5.
The beam size without uvtaper is 0\farcs38$\times$0\farcs30 with a position angle of $-$65.0$^{\circ}$ for the 1.3 mm continuum, and is 0\farcs41$\times$0\farcs31 with a position angle of $-$64.1$^{\circ}$ for the CO 2--1.
The uvtaper for the 1.3 mm continuum was adopted to be 0\farcs9, and that for CO 2--1 to be 0\farcs8.
The beam size and the rms noise level per channel of each molecular line are listed in Table \ref{tab:summary}.

\begin{deluxetable*}{llccccc}
\tablecolumns{7}
\tablecaption{Summary of the image parameters}
\tablewidth{0pc}
\tablehead{\colhead{project code} & \colhead{line/continuum} & \colhead{rest frequency} & \colhead{Synthesized beam}  & \colhead{uv taper} & \colhead{spectral resolution} & \colhead{rms noise level}  \\
 &  & \colhead{(GHz)} & \colhead{(arcsec, deg.)}  & \colhead{(arcsec.)} & \colhead{km s$^{-1}$} & \colhead{(mJy beam$^{-1}$)} 
}
\startdata
2018.1.00302.S &1.3 mm continuum & 218/232 & 0\farcs38$\times$0\farcs30, $-$65.0 & -- & -- & 0.06 \\
 & 1.3 mm continuum & 218/232 &  1\farcs16$\times$1\farcs04, $-$63.4 & 0\farcs9 & -- & 0.15 \\
 & CO 2--1 & 230.538000 & 0\farcs41$\times$0\farcs31, $-$64.1 & -- & 1.47 & 2.8 \\
 & CO 2--1 & 230.538000 & 1\farcs10$\times$0\farcs99, $-$60.9 &  0\farcs8 & 1.47 & 4.0 \\
 & N$_2$D$^+$ 3--2 & 231.32187 & 1\farcs11$\times$0\farcs99, $-$60.7  & 0\farcs8 & 1.46 & 5.0 \\
 & C$^{18}$O 2--1 & 219.5603541 & 1\farcs11$\times$1\farcs00, $-$59.9 & 0\farcs8 & 1.54 & 6.3 \\
 & H$_2$CO 3$_{2,1}$--2$_{2,0}$ & 218.760066 & 1\farcs13$\times$1\farcs02, $-$61.8 & 0\farcs8 & 1.55 & 4.2 \\
 & H$_2$CO 3$_{2,2}$--2$_{2,1}$ & 218.4756320 & 1\farcs12$\times$1\farcs01, $-$62.1 & 0\farcs8 & 1.55 &  4.2 \\
 & H$_2$CO 3$_{0,3}$--2$_{0,2}$ & 218.2221920 & 1\farcs10$\times$1\farcs03, $-$73.9 & 0\farcs8 & 1.55 & 4.1 \\
 & DCN 3--2 & 217.2386307 & 1\farcs11$\times$1\farcs00, $-$59.4 & 0\farcs8 & 1.56 & 4.0 \\
 & SiO 5--4 & 217.1049800 & 1\farcs12$\times$1\farcs00, $-$60.1 & 0\farcs8 & 1.56 & 4.2 \\
 & DCO$^+$ 3--2 & 216.112580 & 1\farcs12$\times$1\farcs01, $-$59.4 & 0\farcs8 & 1.57 & 4.2 \\
2015.1.00341.S & 1.1 mm continuum & 257/273 & 0\farcs17$\times$0\farcs14, +18.3 & -- & -- &0.04 \\
 & 1.3 mm continuum & 217/230.5 & 1\farcs52$\times$0\farcs88, $-$77.2 & -- & -- & 0.40 \\
 & N$_2$D$^+$ 3--2 & 231.32187 & 1\farcs53$\times$0\farcs91, $-$77.9  & -- & 0.10\tablenotemark{*} & 33 \\
 & C$^{18}$O 2--1 & 219.5603541 & 1\farcs58$\times$0\farcs93, $-$77.3 & -- & 0.10\tablenotemark{*} & 30 \\
 & CO 2--1 & 230.538000 & 1\farcs50$\times$0\farcs89, $-$77.9 & -- & 0.32 & 10 
\enddata
\tablenotetext{*}{Spectral resolution of the image data cube.}
\end{deluxetable*}
\label{tab:summary}

\subsection{1.1 mm and 1.3 mm data from the ALMA archive} \label{subsec:N2Dpobs}

We also retrieved the archival ALMA band 6 data of G208-N2 observed in Cycle-3 (\#2015.1.00341.S; PI; S. Takahashi).
The 1.1 mm continuum data have been observed in the full polarization mode with the 12m array.
Two EBs were observed on 2016 September 16 with 38 antennas.
The array configuration was C40-6, which provided the projected baseline range from 12 to 2280 k$\lambda$ and the MRS of 3\farcs0.
Four SPWs with 1.875 GHz bandwidth centered at 256, 258, 272, and 274 GHz provided a total continuum bandwidth of 7.5 GHz.
The phase center of this data set was set to be $\alpha_{\rm ICRS}$ = 5$^{\rm h}$35$^{\rm m}$20.88$^{\rm s}$, $\delta_{\rm ICRS}$ = $-$5\arcdeg 00\arcmin  56\farcs25.
The bandpass calibrator, J0510+1800, and the flux calibrator, J0423-0120, were observed in the first EB.
The phase and polarization calibrators for both EBs are J0541-0211 and J0522-3627, respectively.
The bandpass and flux calibrations of the second EB without bandpass and flux calibrators, have been done using the polarization calibrator, J0522-3627.
The visibility data of each EB were calibrated manually using CASA 4.7.0.
The Stokes {\it I, Q} and {\it U} images were made using the TCLEAN task of CASA 6.4.0.
The Briggs weighting with a robust of +0.5 was adopted.
The resulting beam size is 0\farcs17$\times$0\farcs14 with a position angle of 18.3$^{\circ}$.
In this paper, we only use the Stokes {\it I} image.
The rms noise level of the Stokes {\it I} image is 0.04 mJy beam$^{-1}$.

The 1.3 mm line and continuum observations have been done in the dual polarization mode.
The coordinates of the phase center were the same as those of the 1.1 mm continuum data.
Two SPWs including the N$_2$D$^+$ 3--2 and C$^{18}$O 2--1 lines were set to the high spectral resolution mode with a 35.28 kHz resolution.
The corresponding velocity resolutions were 0.046 km s$^{-1}$ for the N$_2$D$^+$ and 0.048 km s$^{-1}$ for the C$^{18}$O.
The other two SPWs with 468.75 MHz bandwidth centered at 230.5 GHz and 217.1 GHz were used for the continuum.
The CO 2--1 line in the SPW centered at 230.5 GHz was also examined in order to search for the outflow emission.
The spectral resolution of the CO 2--1 line is 244.171 kHz, which corresponds to the velocity resolution of 0.32 km s$^{-1}$.
The images were made using the data obtained with the C36-1 configuration of the 12m array and the ACA 7m array.
The observations with the C36-1 configuration were conducted on 2016 January 29.
The projected baseline range was from 9 to 238 k$\lambda$.
The bandpass calibrator was J0522-3627, which was also used for the flux calibration, and the phase calibrator was J0541-541.
The visibility data has been manually calibrated using CASA 4.7.2.
The 7m array observations were done in 4 EBs on 2016 June 30, July 12, 17, and 19.
The projected baseline range was from 6 to 36 k$\lambda$.
The bandpass, flux, and phase calibrators for the EB of July 12 were J0538-4405, J0854+2006, and J0607-0834, respectively.
Those for the other three EBs were J0522-3627, J0522-3627, and J0542-0913, respectively.
The visibility data were calibrated by the ALMA observatory using CASA 4.6.0.

The calibrated visibility data of the 12m array and ACA were imaged together using the TCLEAN task of CASA 5.5 after continuum subtraction.
The MRS of the combined data is 38\farcs8.
The Briggs weighting with a robust of +0.5 provided the resulting beam size of $\sim$1\farcs5$\times$0\farcs9 for the N$_2$D$^+$, C$^{18}$O, CO and 1.3 mm continuum.
The rms noise level of the 1.3 mm continuum image is $\sim$0.4 mJy beam $^{-1}$.
The N$_2$D$^+$ and C$^{18}$O lines were imaged every 0.1 km s$^{-1}$.
The rms noise level per channel is $\sim$33 mJy beam$^{-1}$ for the N$_2$D$^+$, and is $\sim$30 mJy beam$^{-1}$ for the C$^{18}$O.
The CO line was imaged with the original spectral resolution of 0.32 km s$^{-1}$.
The rms noise level per channel is $\sim$10 mJy beam$^{-1}$.

\section{Results} \label{sec:results}

\subsection{1.3 mm and 1.1 mm continuum -- A compact nucleus in the filamentary cloud} \label{subsec:continuum}

\begin{figure*}
\epsscale{1.0}
\plotone{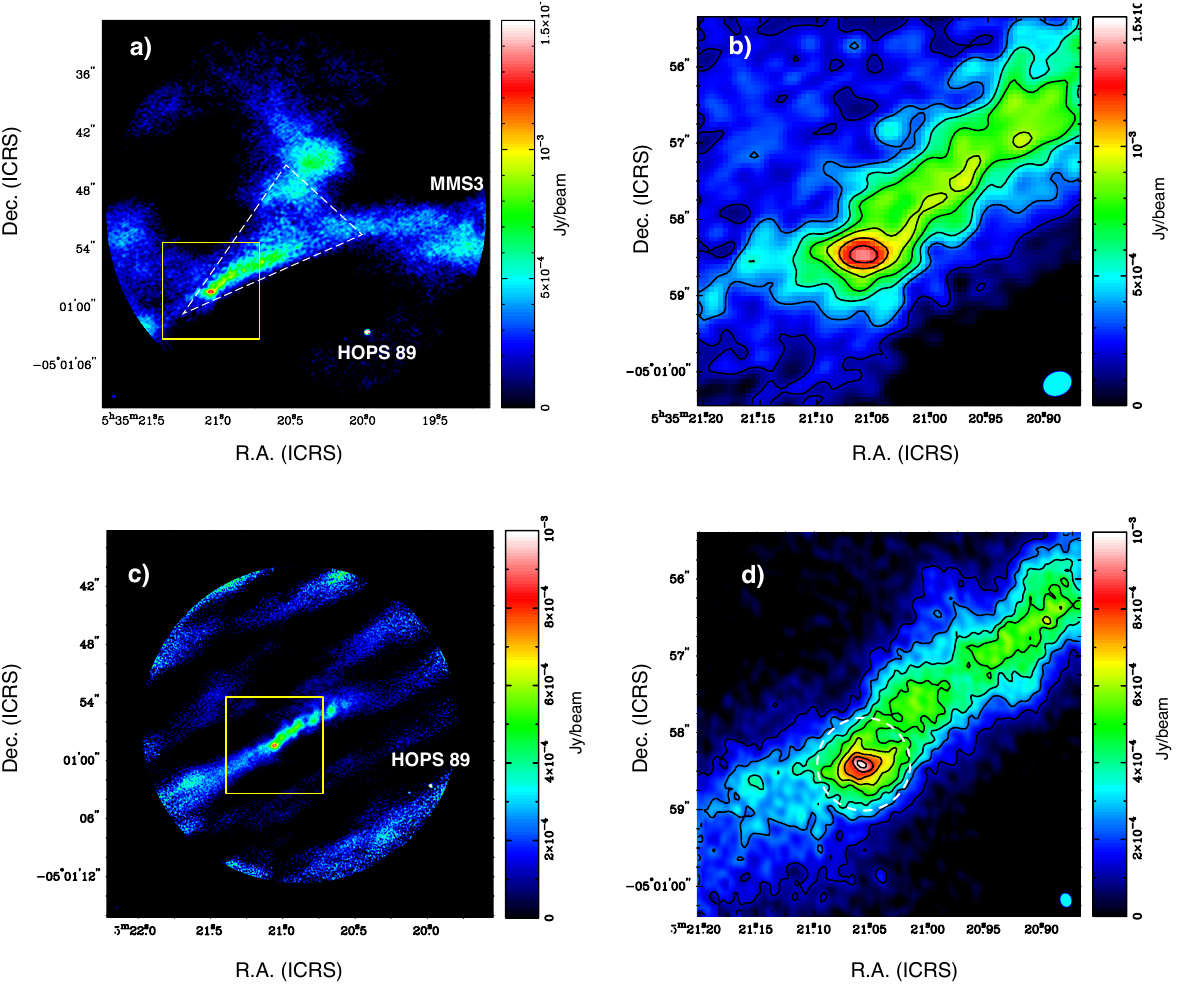}
\caption{a) ALMASOP 1.3 mm continuum emission image obtained using the combined TM1+TM2+ACA data without uvtaper.
Dashed triangle delineates the envelope surrounding the filamentary structure.
The point source in the southwestern part of the field of view is a flat spectrum YSO HOPS 89 \citep{Meg12,Fur16}.
b) A close-up view of the yellow box in panel (a) including the ^^ ^^ nucleus".
Contours are drawn every 3$\sigma$ with the lowest contour of 3$\sigma$.
The 1$\sigma$ level is 0.06 mJy beam$^{-1}$.
The synthesized beam (0\farcs38$\times$0\farcs30 with a P.A. of $-$65.0$^{\circ}$) is drawn as a cyan ellipse in the bottom right corner.
c) 1.1 mm continuum image obtained using the ALMA archival data.
d) A close-up view of the yellow box in panel (c).
Contours are drawn every 3$\sigma$ with the lowest contour of 3$\sigma$.
The 1$\sigma$ level of 0.04 mJy beam$^{-1}$.
The synthesized beam drawn in the bottom right corner has a size of 0\farcs17$\times$0\farcs14 with a P.A. of $+$18.3$^{\circ}$.
Dashed open circle delineates the nucleus area with a radius of 0\farcs6 from the continuum peak.
 \label{fig:cont}}
\end{figure*}

Figure \ref{fig:cont} shows the 1.3 mm and 1.1 mm dust continuum images. 
Figures \ref{fig:cont}a and \ref{fig:cont}b are obtained with the combined TM1+TM2+ACA data of the ALMASOP project, and Figures \ref{fig:cont}c and \ref{fig:cont}d are the Stokes $I$ images of the 1.1 mm archival polarization data.
Figures \ref{fig:cont}a and \ref{fig:cont}c show that the continuum emission from G208-N2 exhibits the narrow filamentary structure with a position angle of $\sim$120$^{\circ}$, which is roughly the same as that of the OMC-3 ridge.
The ALMASOP 1.3 mm image shows that the filament is surrounded by the triangle-shaped envelope, the outer boundary of which is indicated by the dashed triangle in Figure \ref{fig:cont}a, the western extension of which is connected to the adjacent protostellar core, OMC-MMS3.
This extended emission is not sampled in the 1.1 mm data due to the lack of the short spacing data.
The bright point source in the southwestern part of the field of views in both Figures \ref{fig:cont}a and \ref{fig:cont}c is a flat spectrum YSO HOPS 89 \citep{Meg12,Fur16}.
It is obvious that HOPS 89 is not embedded in the dense filament traced by the 1.3 mm and 1.1 mm continuum emission.

To the eastern end of the filamentary structure, there is a compact source (hereafter referred as a ``nucleus'').
The position, peak flux density, integrated flux and size of the nucleus were derived from the 2D gaussian fitting in the image plane.
The fitting results are summarized in Table \ref{tab:nucleus}.
The peak flux densities at 1.3 mm and 1.1 mm correspond to the brightness temperatures of 0.27 K and 0.64 K, respectively, which are much lower than the dust temperature at the center of prestellar cores, 6.5 K \citep{Cra07,Ket10}.
This means that the beam-averaged continuum emission from the nucleus is optically thin at 1.3 mm and 1.1 mm.
The size of the nucleus after deconvolved with the beam is 1\farcs34 (520 au) $\times$ 0\farcs82 (320 au) in the 1.3 mm image and is 0\farcs91 (360 au) $\times$ 0\farcs60 (230 au) in the 1.1 mm image.
The nucleus is elongated along the E-W direction with a position angle of $-$82$^{\circ}$ in both 1.3 mm and 1.1 mm images.
This direction is neither parallel nor perpendicular to the filamentary structure.
The mass of the nucleus is calculated using the formula:
\begin{equation}
    M_{\rm gas}=\frac{F_{\nu} D^2}{\kappa_{\nu}B_{\nu}(T_{\rm dust})},  
    \label{eq:mass}
    \end{equation} 
where $\nu$ is the frequency corresponds to the observed wavelength, $F_{\nu}$ is the total integrated flux,  $D$ is the source distance, $\kappa_{\nu}$ is the dust mass opacity, $T_{\rm dust}$ is the dust temperature, and $B_{\nu}(T_{\rm dust})$ is the Planck function at a temperature of $T_{\rm dust}$. 
We adopt the theoretical dust opacity values of $\kappa_{\rm 230 GHz}=$ 0.011 cm$^2$g$^{-1}$ and $\kappa_{\rm 260 GHz}=$ 0.0145 cm$^2$g$^{-1}$ for the MRN size distribution with thin ice mantles at a number density of 10$^8$ cm$^{-3}$ \citep{Oss94}, and a gas-to-dust ratio of 100.
The dust temperature is assumed to be 11 K, which is the same as the N$_2$D$^+$ excitation temperature  toward the nucleus derived from the hyperfine fitting (see section \ref{subsubsec:fitting}).
The derived mass is $\sim$0.09 $M_{\odot}$.
Assuming that the nucleus is a sphere with an effective radius of 150--200 au, the mean gas density is estimated to be (3.1--8.5)$\times$10$^8$ cm$^{-3}$.

\begin{deluxetable*}{lccccccccccc}
\tablecolumns{12}
\tablewidth{0pc}
\tablecaption{Parameters of the nucleus}
\tablehead{
\colhead{Wavelength}&\multicolumn{2}{c}{Position}&\colhead{}  & \colhead{$S_{\rm peak}$} &  \colhead{$S_{int}$} &\colhead{} &
\multicolumn{3}{c}{Deconvolved size} & \colhead{Mass\tablenotemark{a}} & \colhead{Mean density\tablenotemark{b}}\\
\cline{2-3} \cline{8-10}
\colhead{} &   \colhead{${\alpha}_{\rm ICRS}$} & \colhead{$\delta_{\rm ICRS}$} & \colhead{} & \colhead{mJy beam$^{-1}$} & \colhead{mJy} & \colhead{} &\colhead{major(\arcsec)}   & \colhead{minor(\arcsec)} & \colhead {P.A. ($\arcdeg$)} & \colhead{$M_{\odot}$} & \colhead{$\times$10$^8$ cm$^{-3}$}}
\startdata
1.3 mm & 05 35 21.058 & $-$05 00 58.445 & & 1.25$\pm$0.06 & 13.5 & & 1.34 & 0.82 & $-$82.2 & 0.089 & 3.11\\
1.1 mm & 05 35 21.058 & $-$05 00 58.422 & & 0.89$\pm$0.03 & 21.4 & & 0.91 & 0.60 & $-$82.0 & 0.085 & 8.53\\
\enddata
\tablenotetext{a}{Dust mass opacity values were assumed to be 0.011 cm$^2$g$^{-1}$ for 1.3 mm and 0.0145 cm$^2$g$^{-1}$ for 1.1 mm.}
\tablenotetext{b}{The effective radii are assumed to be 200 au in 1.3 mm and 150 au in 1.1 mm, which are the geometrical means of the semi-major and semi-minor axes.}

\end{deluxetable*}
\label{tab:nucleus}

\subsection{Dense gas tracers\label{subsec:dense_gas}}

\begin{figure*}
\epsscale{1.2}
\plotone{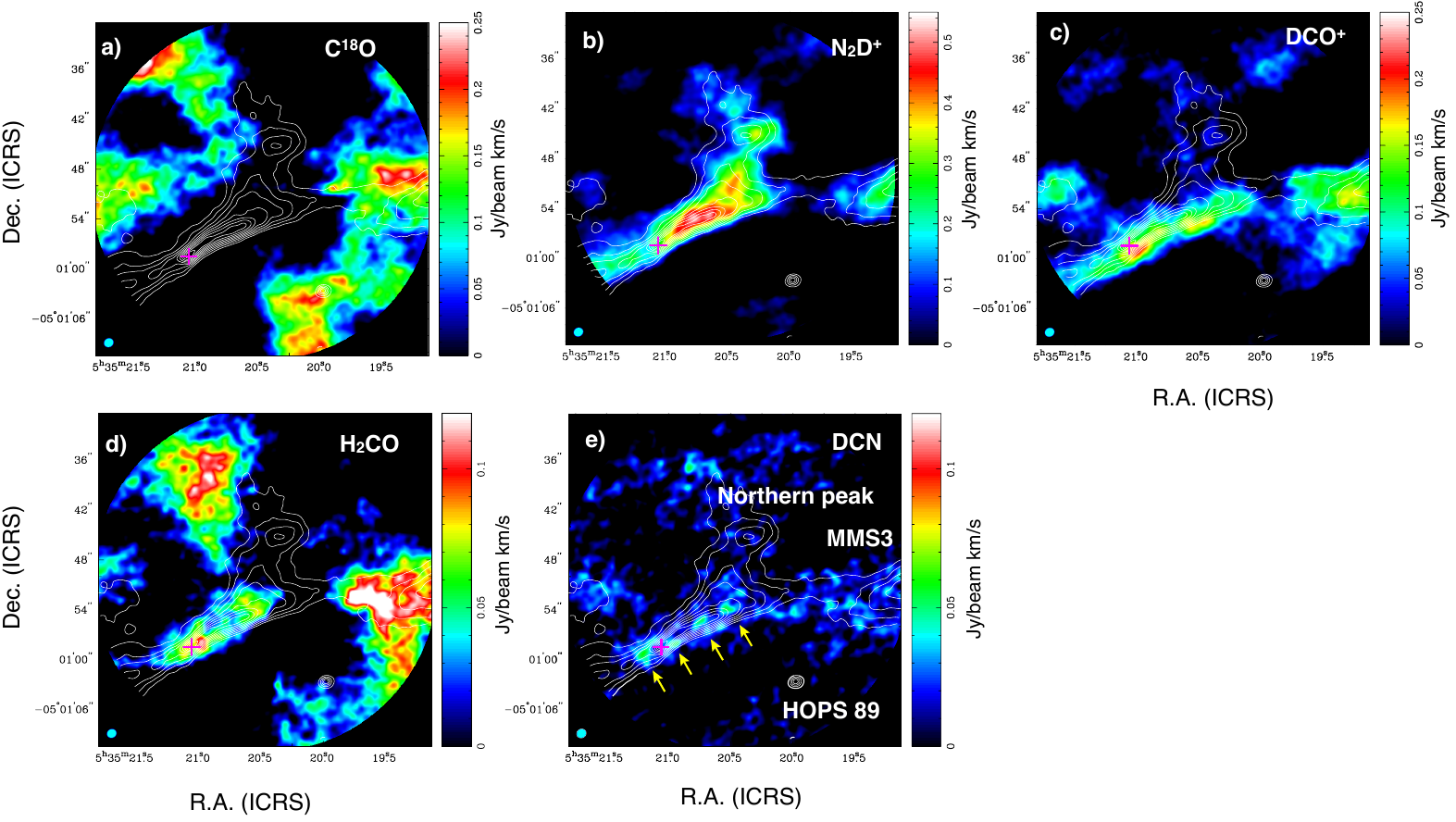}
\caption{Integrated intensity images of a) C$^{18}$O 2--1, b) N$_2$D$^+$ 3--2, c) DCO$^+$ 3--2, d) H$_2$CO 3$_{0,3}$--2$_{0,2}$, e) DCN 3--2 in color overlaid on the tapered 1.3 mm continuum images drawn in contours. 
Contours of the 1.3 mm continuum image are drawn every 10 \% level of the peak flux density.
Magenta cross denotes the position of the nucleus.
Yellow arrows in panel e) indicate the positions of the DCN emission spots. 
\label{fig:lines}}
\end{figure*}

Integrated intensity images of the molecular lines observed with the ALMASOP project are shown in Figure \ref{fig:lines}.
The tapered image of the 1.3 mm continuum emission observed with the ALMASOP is overlaid on each panel.
The tapered continuum image at $\sim$1\arcsec\ spatial resolution reveals that the triangle-shaped envelope surrounding the filament extends to the north and west.
There is a secondary peak to the north of the triangle-shaped envelope (referred to as a ``northern peak'').
The continuum enhancement at the western edge of the field of view is the neighbouring protostellar core MMS3 \citep{Mor21}.
The area enclosed by the 20\% level contour has a size of 17\farcs5 (6,800 au) $\times$ 3\farcs5 (1,400 au) with an aspect ratio of $\sim$5, and that within the 50\% contour level has a size 12\arcsec (4,700 au) $\times$ 1\farcs5 (590 au) with an aspect ratio of $\sim$8.
The area within the 50\% contour level is referred to as a ``continuum filament''.

The C$^{18}$O emission is distributed in the northeast and southwest of the field of view.
The C$^{18}$O emission extending to the west belongs to the neighbouring core, MMS3.
It is remarkable that the C$^{18}$O
 emission is completely missing in the region traced by the continuum including the continuum filament, the triangle-shaped envelope, and the northern peak.
In addition, there is no C$^{18}$O emission in the vicinity of the nucleus.
In contrast to the C$^{18}$O, the N$_2$D$^+$ shows a similar distribution as the continuum emission, as reported by \citet{Sah21}.
The N$_2$D$^+$ emission also comes from the northern peak and MMS3.
The anticorrelation between the C$^{18}$O and N$_2$D$^+$ is consistent with the picture that G208-N2 is an evolved starless core with CO depletion through freeze-out and high deuterium fractionation.
It is also in contrast with the neighboring protostellar core, MMS 5, in which the centrally condensed structure surrounding the protostar is traced well with  C$^{18}$O and barely seen in  N$_2$D$^+$ \citep{Mat19}.
Although the spatial distribution of N$_2$D$^+$ correlates well with that of the dust continuum, the  N$_2$D$^+$ emission is brighter in the western part, and is not enhanced at the nucleus.

The DCO$^+$ emission shows an elongated feature running parallel to the continuum filament. 
The DCO$^+$ ridge is systematically offset by $\sim$2\arcsec\ southward from the continuum filament, and delineates the southern edge of the triangle-shaped evelope.
The DCO$^+$ emission peaks to the south of the nucleus.
The DCO$^+$ emission is also associated with MMS3, but not with the northern peak.

The H$_2$CO emission is detected above 3 $\sigma$ only in the 3$_{0,3}$--2$_{0,2}$ transition with $E_{\rm up}$ = 20.96 K.
The spatially extended H$_2$CO emission is seen in the northern and western parts of the field of view.
The H$_2$CO emission is bright in MMS3, while it is not detected in the northern peak.
The H$_2$CO emission also comes from the region of the continuum filament.
The intensity distribution is an oval-shaped ring surrounding the continuum filament.
Such a ring-like distribution in the H$_2$CO emission is also observed in the well-studied prestellar core, L1544 \citep{Cha19}.
The H$_2$CO does not peak toward the nucleus, but shows two local peaks in the vicinity of the nucleus. 
The brighter one at $\sim$1\arcsec\ west, and the fainter one at $\sim$0\farcs5 east of the nucleus  (the detailed distribution of H$_2$CO in the vicinity of the nucleus is described in section \ref{subsec:heating}). 
The locations of the two emission peaks are close to but not coincide with that of the nucleus.
The other transitions, 3$_{2,2}$--2$_{2,1}$ and 3$_{2,1}$--2$_{2,0}$, with higher energy levels of $E_{\rm up}$ = 68 K were not detected above the 3 $\sigma$ level.

DCN shows four emission spots at 5 $\sigma$ level along the continuum filament.
The positions of the emission spots are indicated by the  yellow arrows in Figure \ref{fig:lines}e.
The DCN emission, the maximum integrated flux density of which is $\sim$0.06 Jy beam$^{-1}$ km s$^{-1}$, is much fainter than those of the other Deuterated molecules, DCO$^+$ ($\sim$0.20 Jy beam$^{-1}$ km s$^{-1}$) and N$_2$D$^+$ ($\sim$0.52 Jy beam$^{-1}$ km s$^{-1}$).
This is probably because of the low gas temperature in G208-N2.
In contrast to the N$_2$D$^+$ and DCO$^+$, which are formed by H$_2$D$^+$ in the temperature range of $<$30 K \citep{Mil89}, DCN is predominantly formed from CH$_2$D$^+$ in the warmer temperature range of 30--70 K \citep{Mil89,Tur01}.
DCN is not entirely absent, because a smaller fraction (22 \%) of DCN can be formed from H$_2$D$^+$ in low temperature \citep{Tur01}.

\subsection{Outflow tracers\label{subsec:CO}}

\begin{figure*}
\epsscale{0.9}
\plotone{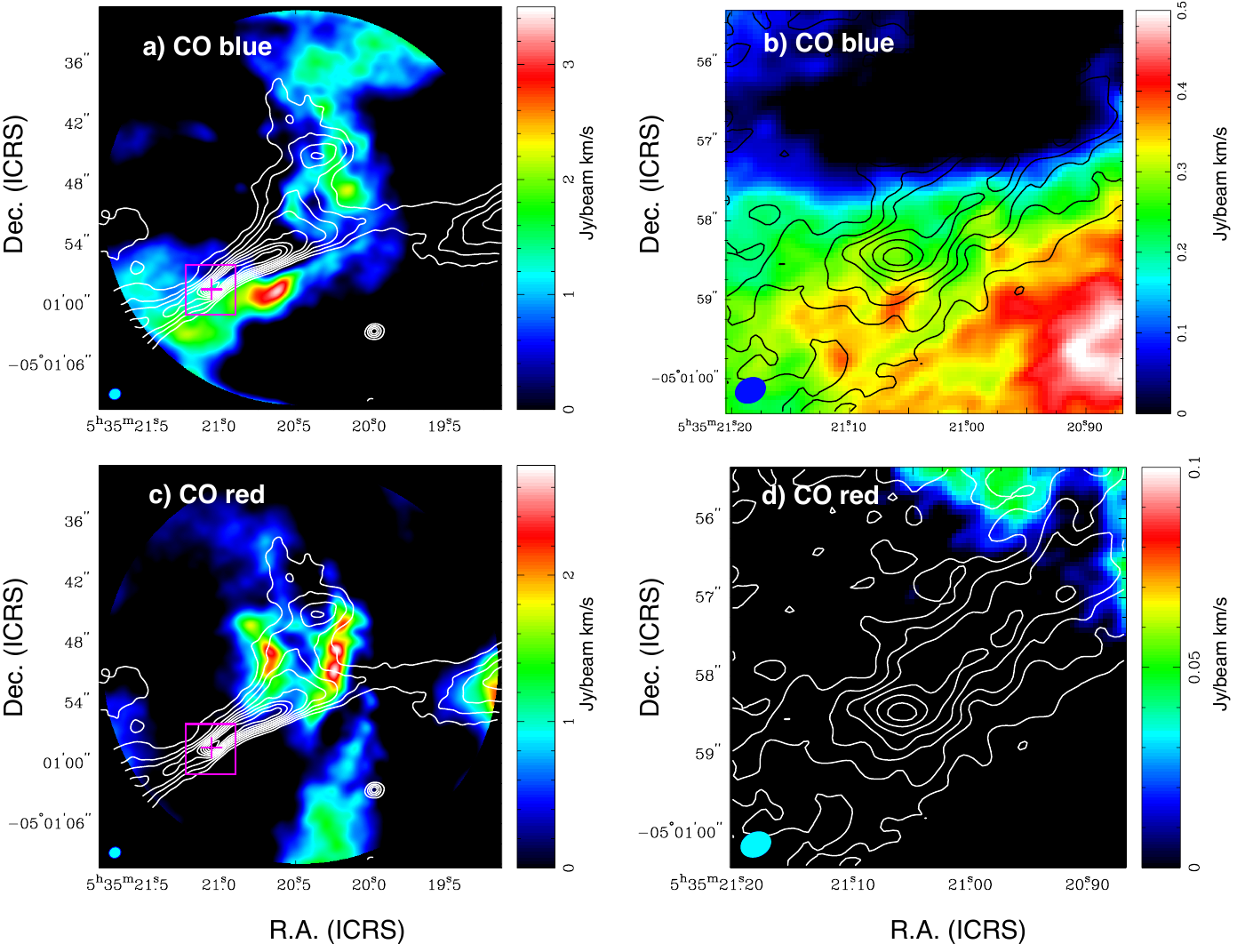}
\caption{Integrated intensity images of CO 2--1 (color) overlaid on the 1.3 mm continuum images drawn in contours. 
The velocity ranges of the blueshifted emission (panels a and b) is $v_{\rm LSR} =$ 0.17--10.33 km s$^{-1}$, and that of the redshifted emission (panels c and d) is $v_{\rm LSR} =$ 12.23--16.04 km s$^{-1}$. 
Panels a) and c) show the tapered CO and continuum images at $\sim$1\arcsec resolution.
Contours of the 1.3 mm continuum image are drawn every 10\% of the peak flux density.
Magenta cross and box denote the position of the nucleus and the area of the close-up view (panels b and d), respectively.
Panels b) and d) show the images without uvtaper in the small region around the nucleus.
Contours of the 1.3 mm continuum image are drawn every 3 $\sigma$ with the 1 $\sigma$ level of 0.084 mJy beam$^{-1}$.}
\label{fig:CO}
\end{figure*}

In order to search for any outflow associated with the nucleus, the spatial distribution of the CO 2--1 emission was examined.
Figure \ref{fig:CO} presents the spatial distributions of the blue- and redshifted CO emission observed with ALMASOP.
The left panels of Figure \ref{fig:CO} (a and c) are the tapered images at $\sim$1\arcsec\ resolution, and the right panels (b and d) are the closeup view of the images without uv taper ($\sim$0\farcs4 resolution). 
The blueshifted CO emission extends to the south and west of the triangle-shaped continuum emission, while the redshifted CO is bright in the two elongated features along the north-south direction.
As shown in Figure \ref{fig:CO}d, there is no redshifted CO emission in the vicinity of the nucleus.
On the other hand, the spatially extended blueshifted CO emission is observed in the eastern part of G208-N2 including the nucleus.
However, this blueshifted emission is not localized to the nucleus.

The CO emission was also examined using the archival data with higher spectral resolution of 0.32 km s$^{-1}$.
The channel map of CO is shown in Appendix \ref{sec:app_chmap}.
Although the CO emission in the velocity range from $V_{\rm LSR}{\sim}$ 6 -- 7.5 km s$^{-1}$ extends to the south of the nucleus, this component is likely to be the part of the extended CO emission along the line of sight rather than the outflow from the nucleus. 

We also examined the SiO emission, which often traces the shocks associated with outflows, and confirmed that there is no SiO emission in the field of view of G208-N2.
Thus, we find no direct evidence of an outflow emanating from the nucleus.

\subsection{N$_2$D$^+$ emission with high spectral resolution\label{subsec:n2dp}}

In order to study the kinematics and physical properties of the dense gas, the high spectral resolution data of N$_2$D$^+$ were examined.
The channel map of N$_2$D$^+$ at 0.1 km s$^{-1}$ resolution and that compared with the channel map of C$^{18}$O are shown in  \ref{fig:chmap} and \ref{fig:N2Dp_C18O}, respectively, in Appendix.
The channel map of C$^{18}$O (Figure \ref{fig:N2Dp_C18O}) confirms that 
the C$^{18}$O emission avoids the region of the filament in all velocity channels.
In the velocity range of 11.4--11.8 km s$^{-1}$, there is a faint C$^{18}$O emission to the south of the nucleus.

\subsubsection{Hyperfine Spectral Fitting \label{subsubsec:fitting}}

Since the N$_2$D$^+$ 3--2 line consists of numerous hyperfine components, the line parameters such as LSR velocity ($V_{\rm LSR}$), line width (${\Delta}V_{\rm FWHM}$), optical depth ($\tau$), and excitation temperature ($T_{\rm ex}$) can be determined from hyperfine fitting.
All the hyperfine components are assumed to have the same excitation temperature and line width.
The rest frequency and relative intensity of each hyperfine component were referred to the CDMS catalog \citep{Mul01,Mul05}. 

\begin{figure*}
\epsscale{1.15}
\plotone{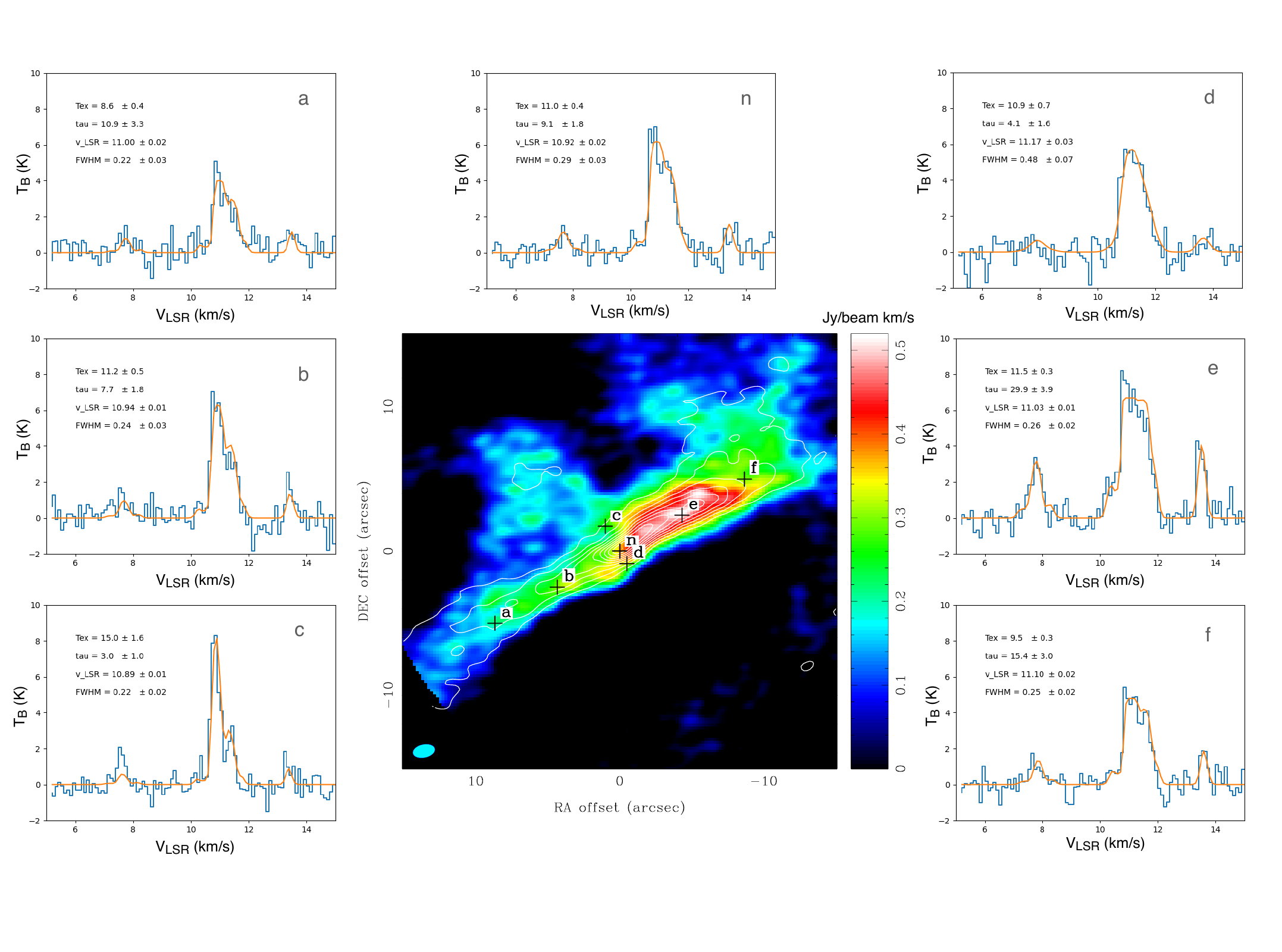}
\caption{N$_2$D$^+$ 3--2 spectra at seven representative positions marked in the mom 0 image in color.
Red curves are the results of the hyperfine spectral fitting.
The parameters derived from the fitting, excitation temperature, total optical depth, LSR veloity, and the line width in FWHM, are given in each panel.
1.3 mm continuum intensity is overlaid on the mom 0 image as white contours.
Contours are drawn every 10 \% level of the peak intensity.
The origin of the map is the position of the nucleus (marked as ^^ ^^ n").
\label{fig:spectra}}
\end{figure*}

Figure \ref{fig:spectra} shows the line profiles and the fitting results toward the nucleus and several representative positions along the major and minor axes of the filament through the nucleus.
The N$_2$D$^+$ 3--2 line consists of three groups of hyperfine (hf) components, low-V hf group (7.4--8.0 km s$^{-1}$), main-V hf group (10.2--12.0 km s$^{-1}$), and high-V hf group (13.2--13.7 km s$^{-1}$).
The low-V and high-V hf groups, which are often referred to as the satellite hf groups, are clearly seen in most of the region.
If the relative strengths of the hf components follow statistical equilibrium, the low-V and high-V hf groups only account for 3.37 \% and 3.73 \%, respectively, of the total line strength.
The observed integrated intensity of each hf group, however, is $\sim$10 \% of the total integrated intensity.
This implies that the optical depth of the N$_2$D$^+$ line is significant, especially in the western part of the continuum filament.
Although the bright satellite hf groups could be attributed to non-LTE excitation anomalies as in the case of N$_2$H$^+$ \citep{dan06}, such a non-LTE excitation becomes significant for the case of large optical depth.
The line parameters toward the nucleus derived from the hf spectral fitting are $T_{\rm ex}$ = 11.0$\pm$0.4 K, ${\tau}_{\rm tot}$ = 9.1$\pm$1.8, $V_{\rm LSR}$ = 10.92$\pm$0.02 km s$^{-1}$, and ${\Delta}V_{\rm FWHM}$ = 0.29$\pm$0.03 km s$^{-1}$.

\begin{figure*}
\plotone{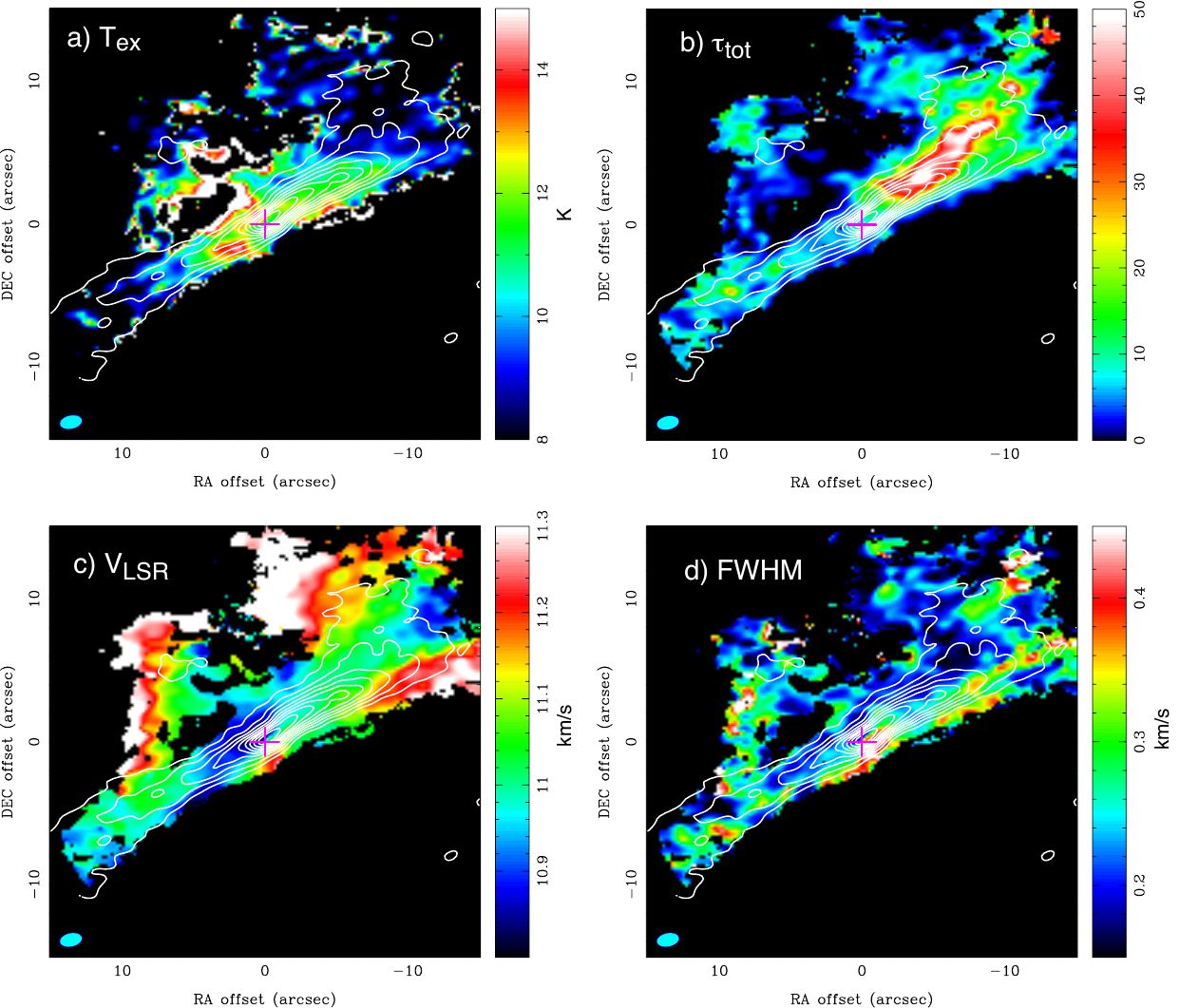}
\caption{Line parameters for N$_2$D$^+$ 3--2 derived from the hyperfine fitting in color, a) excitation temperature, b) total optical depth, c) LSR velocity, and d) line width.
1.3 mm continuum emission observed in 2015.1.00341.S is overlaid as contours.
Contours are drawn every 1.563 mJy beam$^{-1}$, which corresponds to the 10 \% level of the peak intensity.
The origin of the map is the position of the nucleus (magenta cross).
\label{fig:N2Dp_parameters}}
\end{figure*}

The spatial distributions of the excitation temperature, optical depth, LSR velocity, and line width derived from the hf spectral fitting are presented in Figure \ref{fig:N2Dp_parameters}.
The excitation temperature $T_{\rm ex}$ ranges from $\sim$8 to $\sim$15 K with the average value of $\sim$11 K (Figure \ref{fig:N2Dp_parameters}a).
The excitation temperature remains relatively stable at around 11-12 K within the continuum filament encompassing the nucleus, but experiences a slight increase by a few K in the northeastern, southeastern, and southern peripheries.
The enhanced excitation temperatures observed in these peripheral areas are likely attributed to external heating influences.
As shown in the C$^{18}$O channel map (Figure \ref{fig:N2Dp_C18O}), the diffuse C$^{18}$O emission is extended in the northeastern region of the filamentary cloud, suggesting the presence of gas with $>$ 20 K in this area.
The northeastern periphery of the filamentary cloud corresponds to the area where the dense gas is facing the diffuse warmer gas traced by the C$^{18}$O emission.
The elevated excitation temperatures observed in the southeastern and southern peripheries are likely the result of external pressure  from the south (see section \ref{subsubsec:kinematics}).

Figure \ref{fig:N2Dp_parameters}b shows that the optical depth of the N$_2$D$^+$ emission $\tau_{\rm tot}$ is extremely large ($>$10) in the  filament traced by the continuum emission; especially, $\tau_{\rm tot}$ is higher than 30 in the northwestern part of the filament.
This is partly because of the calculation error due to the non-gaussian line profiles with broad redshifted wings that mimic the flat-topped profiles of the optically thick lines (e.g. Figure \ref{fig:spectra}e ). 
However, the bright satellite hf groups suggest that the optical depth of the line is significant in the filament including the nucleus.

The LSR velocity (Figure \ref{fig:N2Dp_parameters}c) shows the velocity gradient along the NE-SW direction, which is along the minor axis of G208-N2.
The overall velocity gradient seen in the N$_2$D$^+$ is consistent with that of the large scale single-dish N$_2$H$^+$ 1--0 data, in which redshifted emission with $>$ 11.2 km s$^{-1}$ appears to the northeastern and southwestern sides of G208-N2 with a velocity gradient along its minor axis \citep{Luo22}.
Note that the redshifted velocity ($>$ 11.3 km s$^{-1}$) at the southern edge is due to the redshifted wing shown in Figure {\ref{fig:spectra}}.
The velocity also varies along the major axis; the LSR velocity of $\sim$11.0--11.1 km s$^{-1}$ in the northwestern and southeastern regions decreases to the most blueshifted velocity of $\sim$10.8 km s$^{-1}$ in the vicinity of the nucleus.
In addition, the LSR velocity changes significantly across the nucleus from $V_{\rm LSR}{\sim}$ 10.8 km s$^{-1}$ in the northeast to $\sim$ 11.2 km s$^{-1}$ in the southwest.

 The line width (Figure \ref{fig:N2Dp_parameters}d) tends to be enhanced ($>$ 0.3 km s$^{-1}$) along the southern edge of the continuum filament due to the redshifted wing emission.
 On the other hand, it is rather narrow ($<$ 0.25 km s$^{-1}$) in the northern part of the continuum filament where the effect of the redshifted wing emission is negligible.
 The nonthermal velocity dispersion ${\Delta}v_{\rm nt}$ is derived by
\begin{equation}
    {\Delta}v_{\rm nt} = \left[\frac{({\Delta}v'_{\rm FWHM})^2}{8\ln2}-\frac{kT_{\rm kin}}{{\mu}(N_2D^+)}\right]^{1/2},
    \label{eq:nonthermal}
\end{equation}
where ${\Delta}v'_{\rm FWHM}$ is the line width obtained from the hf spectral fitting after the correction of the spectral resolution.
Here, $k$ is the Boltzman constant, $T_{\rm kin}$ is the gas kinetic temperature, and $\mu$ is the molecular mass.
 Assuming that the gas kinetic temperature is $\sim$11 K, the non-thermal velocity dispersion is calculated to be $<$ 0.08 km s$^{-1}$, which is smaller than the sound speed, 
 \begin{equation}
     c_{\rm s} = \left[ \frac{kT_{\rm kin}}{{\mu_p}m_H}\right]^{1/2}
     \label{eq:thermal}, 
 \end{equation}
 where $\mu_p$ is the mean molecular mass per free particle, 2.37 and $m_H$ is the mass of atomic hydrogen.
 Using the same temperature, $c_s$ is 0.20 km s$^{-1}$.
 This implies that the turbulent velocity of the gas in the continuum filament is subsonic.

\subsubsection{Kinematics of the dense gas\label{subsubsec:kinematics}}

\begin{figure*}
\plotone{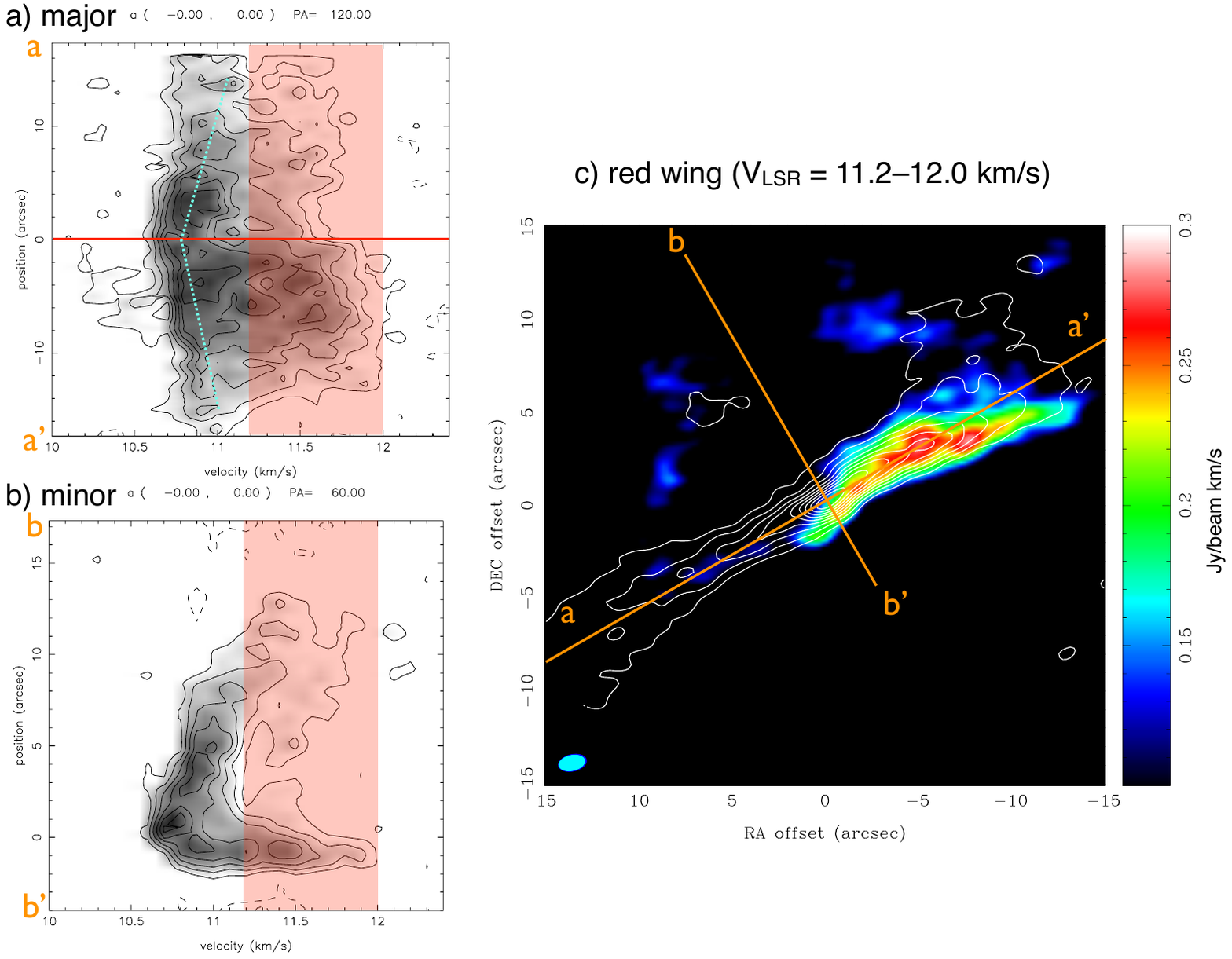}
\caption{Position-Velocity diagrams of N$_2$D$^+$ along the a) major (a--a') and b) minor (b--b') axes of the filament through the nucleus.
The orange horizontal line and shaded area in each panel denote the location of the nucleus and the velocity range of the red wing ($V_{\rm LSR}$ = 11.2--12.0 km s$^{-1}$), respectively.
The cyan dotted lines in panel a) represent the velocity gradients along the major axis.
c) Red wing of the N$_2$D$^+$ in color overlaid on the 1.3 mm continuum in white contours.
Orange lines (a--a' and b--b') are the cuts of the P-V diagrams.
}
\label{fig:N2Dp_PV}
\end{figure*}

Figures \ref{fig:N2Dp_PV}a and b show the Position-Velocity (PV) diagrams of N$_2$D$^+$ along the major and minor axes, respectively, of the filament through the nucleus.
Figure \ref{fig:N2Dp_PV}a reveals that the N$_2$D$^+$ line intensity drops sharply on the blue side, while it decreases gradually on the red side.
The enhancement of the wing on the red side is significant in the western part of the filament.
There are  small velocity gradients that decrease toward the nucleus from both sides.
The velocity gradients are $\sim$0.3 km s$^{-1}$ and $\sim$0.2 km s$^{-1}$ per 12\arcsec ~in the eastern side and the western side, respectively (the cyan lines in Figure \ref{fig:N2Dp_PV}a). 
These velocity gradients correspond to 13.2 km s$^{-1}$ pc$^{-1}$ and 8.8 km s$^{-1}$ pc$^{-1}$, respectively.
On the other hand, the velocity structure along the minor axis (Figure \ref{fig:N2Dp_PV}b) exhibits a pronounced V-shaped pattern.
The velocity gradually decreases from $V_{\rm LSR}{\sim}$ 11.27 km s$^{-1}$ at 10\arcsec\ north to ${\sim}$10.73 km s$^{-1}$ at the nucleus.
This velocity gradient corresponds to 28.5 km s$^{-1}$ pc$^{-1}$.
Then, the broad redshifted wing appears suddenly at the southern edge, toward which the line intensity drops sharply.
This implies that the dense gas is compressed by the external pressure from the southwest.
Figure \ref{fig:N2Dp_PV}c shows that the redshifted wing is extended over the filament, suggesting that the gas in the filament is highly perturbed by the external pressure.
The nucleus is located at the boundary of the perturbed and quiescent regions.
The significant velocity change across the nucleus shown in Figure \ref{fig:N2Dp_parameters}c is due to this redshifted perturbed component.

\section{Analysis\label{sec:analysis}}

\subsection{H$_2$ column density\label{subsec:H2_column_density}}

The column density of molecular hydrogen can be estimated using the 1.3 mm continuum emission by the formula:
\begin{equation}
  N(\rm{H}_2)=\frac{S_{\nu}}{{\Omega_{\rm m}}{\mu}m_{H}\kappa_{\nu}B_{\nu}(T_{\rm dust})},  
\end{equation}
where $S_{\nu}$ is the flux density, $\Omega_{\rm m}$ is the solid angle of the beam, $\mu$ is the mean molecular weight, i.e. 2.8, $m_H$ is the mass of the atomic hydrogen, $\kappa_{\nu}$ is the dust mass opacity, and $B_{\nu}(T_{\rm dust})$ is the Planck function at a temperature of $T_{\rm dust}$. 
We adopt the theoretical dust opacity of $\kappa_{\rm 230 GHz}=$ 0.0106 cm$^2$g$^{-1}$ for the MRN size distribution with thin ice mantles at a number density of 10$^7$ cm$^{-3}$ \citep{Oss94}.
The gas to dust ratio is assumed to be 100.
Here, the dust temperature is assumed to be same as the excitation temperature of N$_2$D$^+$ derived from the hyperfine fitting.
Since the excitation temperature of N$_2$D$^+$ agrees well with the kinetic temperature derived from NH$_3$ \citep{Li13}, this assumption should be reasonable for the region where the gas and dust temperatures are coupled.

The derived H$_2$ column density distribution is shown in Figure \ref{fig:column_density}a.
Since 1.3 mm continuum emission is optically thin, the H$_2$ column density distribution follows the intensity distribution of the continuum emission.
The H$_2$ column density exceeds 5$\times$10$^{23}$ cm$^{-2}$ in the area of the continuum filament.
The mass of the continuum filament (including the nucleus) is 0.59 $M_{\odot}$.
Assuming the cylindrical geometry with a diameter of 1\farcs5 (590 au), the mean density of the continuum filament is calculated to be 5.9$\times$10$^{7}$ cm$^{-3}$.
The H$_2$ column density toward the position of the nucleus is estimated to be (8.4$\pm$0.2)$\times$10$^{23}$ cm$^{-2}$.

\subsection{N$_2$D$^+$ Column Density\label{subsec:N2Dp_column_density}}

The column densities of N$_2$D$^+$ have been calculated using the excitation temperatures ($T_{\rm ex}$) and optical depths ($\tau_{\rm tot}$) derived from the hf fitting.
The N$_2$D$^+$ emission is optically thick in the entire area of G208-N2, especially $\tau_{\rm tot}$ is larger than 10 in the filament including the nucleus.
Therefore, the column density was estimated using the integrated intensities of the low-V and high-V hf groups following the method described in \citet{bou12}.
Most of the optical depth of the N$_2$D$^+$ line is due to the main-V hf group, the contribution of which is 0.929 to the total line intensity (normalized to 1.0).
On the other hand, the contribution of the low-V and high-V hf groups to the total line intensity are 0.0373 and 0.0337, respectively.
Therefore, the emission of these satellite groups is considered to be optically thin except for the area with extremely large $\tau_{\rm tot}$ of $>$ 30.
We used the integrated intensities of the low-V (7.4--8.0 km s$^{-1}$) and high-V (13.2--13.7 km s$^{-1}$) hf groups and assumed that the emission is optically thin.
Then, the derived column density was scaled by the inverse of the relative line intensity, 1/(0.0373+0.0337).

\begin{figure*}
\epsscale{1.17}
\plotone{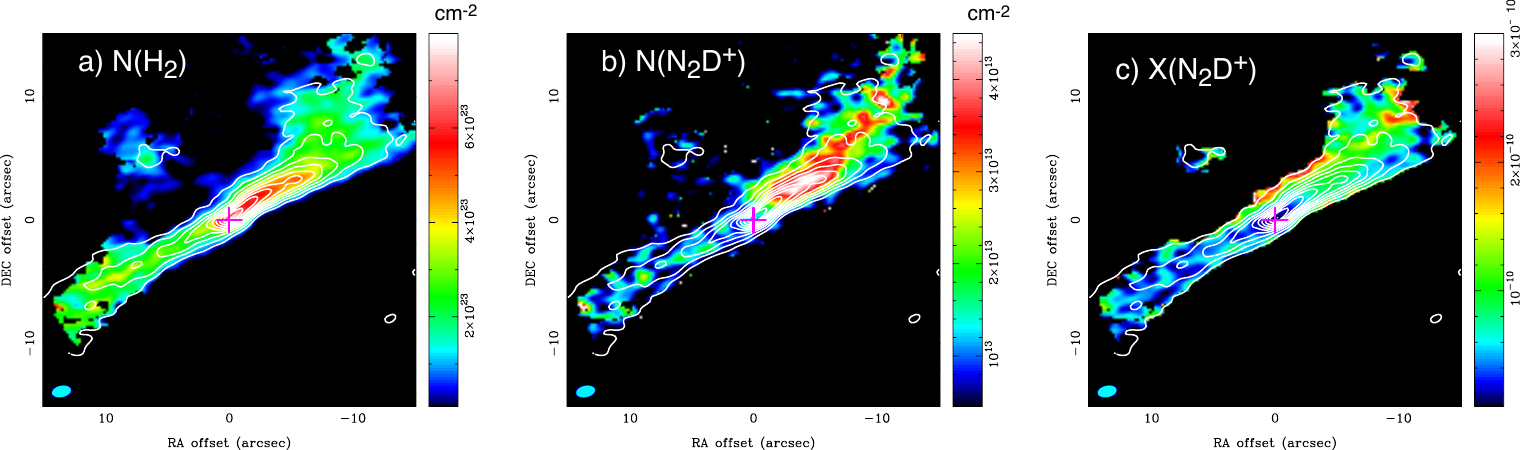}
\caption{a) H$_2$ column density derived from the 1.3 mm continuum emission, assuming the dust temperature same as the excitation temperature of N$_2$D$^+$.
The range of the color scale is from 1.0 $\times$ 10$^{22}$ to 8.0 $\times$ 10$^{23}$ cm$^{-2}$.
b) Column density of N$_2$D$^+$ derived from the integrated intensity of the satellite hf groups scaled by the inverse of the relative line intensity, 1/(0.0373+0.0337).
The range of the color scale is from 4.5 $\times$ 10$^{12}$ to 4.5 $\times$ 10$^{13}$ cm$^{-2}$.
c) Fractional abundance of N$_2$D$^+$ derived from column densities of H$_2$ (a) and N$_2$D$^+$ (b).
The range of the color scale is from 1.0 $\times$ 10$^{-11}$ to 3.0 $\times$ 10$^{-10}$.
1.3 mm continuum emission observed in 2015.1.00341.S is overlaid as white contours.
Contours are drawn every 1.563 mJy beam$^{-1}$, which corresponds to the 10 \% level of the peak intensity.
The origin of the map is the position of the nucleus (magenta cross).
\label{fig:column_density}}
\end{figure*}

The column density of N$_2$D$^+$, $N$(N$_2$D$^+$), derived in this method is shown Figure \ref{fig:column_density}b.
The $N$(N$_2$D$^+$) is higher than 5$\times$10$^{12}$  cm$^{-2}$ in the entire region of G208-N2, with  values higher than 2$\times$10$^{13}$ cm$^{-2}$ in the northwestern part with large optical depth.
The N$_2$D$^+$ column density values derived here are consistent with (7.5$\pm$0.7)$\times$10$^{12}$ cm$^{-2}$ derived from the $J$=1--0 transition observed with the NRO 45m telescope \citep{Kim20}.
The $N$(N$_2$D$^+$) toward the nucleus is estimated to be (1.3$\pm$0.3)$\times$10$^{13}$ cm$^{-2}$.
Here, the error on $N$(N$_2$D$^+$) is estimated from the rms noise level of the integrated intensity \citep{Cas02}.
The highest value of $\sim$4.5$\times$10$^{13}$ cm$^{-2}$ is obtained at $\sim$4\farcs3 northwest of the nucleus.

The $N$(N$_2$D$^+$) values derived from the satellite hf groups are compared with those derived from the total integrated intensities (including the main-V hf group) and total optical depths, $\tau_{\rm tot}$, in the eastern part with $\tau_{\rm tot} <$ 10. 
The values derived from the satellite groups are lower by a factor of $\sim$2 as compared to those derived from the total integrated intensity and $\tau_{\rm tot}$. 
This implies that the $N$(N$_2$D$^+$) shown in Figure \ref{fig:column_density}b could be underestimated, even though the derived values of $>$2$\times$10$^{13}$ cm$^{-2}$ are $\sim$10 times higher than the N$_2$D$^+$ column densities measured in the nearby low-mass star forming regions such as $\rho$-Oph A-N6 \citep{bou12} and B1-bN/B1-BS \citep{hua13}.

\subsection{N$_2$D$^+$ abundance\label{subsec:N2Dp_abundance}}

The fractional abundance of N$_2$D$^+$, $X$(N$_2$D$^+$) was derived from the column densities of N$_2$D$^+$ and H$_2$, and presented in Figure \ref{fig:column_density}c.
The $X$(N$_2$D$^+$) is enhanced to $>$10$^{-10}$ in the northwestern part because of the high $N$(N$_2$D$^+$) of $>$ 2$\times$10$^{13}$ cm$^{-2}$ and moderate $N$(H$_2$) of $\sim$2$\times$10$^{23}$ cm$^{-2}$.
The $X$(N$_2$D$^+$) is $\sim$8$\times$10$^{-11}$ in the western part of the filament, and decreases to $<$2$\times$10$^{-11}$ in the vicinity of the nucleus, toward which the N$_2$D$^+$ abundance shows the local minimum.
The $X$(N$_2$D$^+$) at the position of the nucleus is derived to be (1.6$\pm$0.4)$\times$10$^{-11}$ cm$^{-2}$, which is one order of magnitude lower than the highest value in the northwestern region.

\section{Physical properties of the filament and nucleus \label{sec:filament and nucleus}}

\subsection{Physical properties of the filament \label{subsec:filament}}

Using the mean density of the continuum filament derived in section \ref{subsec:H2_column_density}, the line mass of the continuum filament, $M_{\rm line}$, is calculated to be $M_{\rm line}{\sim}$1.67${\times}$10$^{16}$ g cm$^{-1}$ (= 5.9 $M_{\odot}$ pc$^{-1}$).
On the other hand, the critical line mass for an infinite filament in hydrostatic equilibrium is calculated by \citep{Ost64,Sto63}
 \begin{equation}
     M_{\rm crit}= \frac{2{\Delta}v_{\rm tot}^2}{G},
 \end{equation}
 where ${\Delta}v_{\rm tot}$ is a total velocity dispersion including both thermal and nonthermal contributions, and G is the gravitational constant.
 The total velocity dispersion, ${\Delta}v_{\rm tot}$ is
\begin{equation}
    {\Delta}v_{\rm tot} = \left[ {\Delta}v_{\rm nt}^2 + c_s^2\right]^{1/2}.
    \label{eq:dv_tot}
\end{equation}
The typical line width, ${\Delta}v_{\rm FWHM}$, measured in the northern part of the continuum filament with less affected by the redshifted wing is $\sim$0.25 km s$^{-1}$.
The nonthermal dispersion calculated by Equation (\ref{eq:nonthermal}) is 0.08 km s$^{-1}$, and the total velocity dispersion calculated by Equation (\ref{eq:dv_tot}) is 0.21 km s$^{-1}$ for $T_{\rm kin}$ = 11 K.
The critical line mass is calculated to be $M_{\rm crit}=1.3{\times}10^{16}$ g cm$^{-1}$ (= 20.5 $M_{\odot}$ pc$^{-1}$).
The $M_{\rm line}$/$M_{\rm crit}$ is $\sim$1.26, implying that the filament is close to the critical state between stable and unstable.

If magnetic field is taken into account, the critical line mass can be larger.
The critical line mass including the contribution of the magnetic field can be calculated using the formula described in \citet{Li22a}.
The magnetic field strength in the dense clump including G208-N2 was estimated to be $\sim$0.78 mG assuming that the OMC-3 ridge lies close to the plane of the sky \citep{Li22b}.
If the magnetic field strength in the continuum filament is comparable to that in the larger scale clump, i.e. the filament is formed through the contraction along the magnetic field that is roughly perpendicular to the OMC-3 ridge \citep{Mat01,Hou04,Li22b}, the contribution of the magnetic field to the critical line mass is negligible.
If the magnetic field strength is comparable to that observed in the protostellar envelopes, i.e. a few to 5 mG \citep{Aso21,Gir06,Hul17}, the critical line mass become larger by a factor of $\sim$1.5.
These imply that the magnetic field does not affect the dynamical state of the continuum filament significantly.

\subsection{{Radial density profile of the nucleus} \label{subsec: density profile}}

\begin{figure*}
\epsscale{1.1}
\plotone{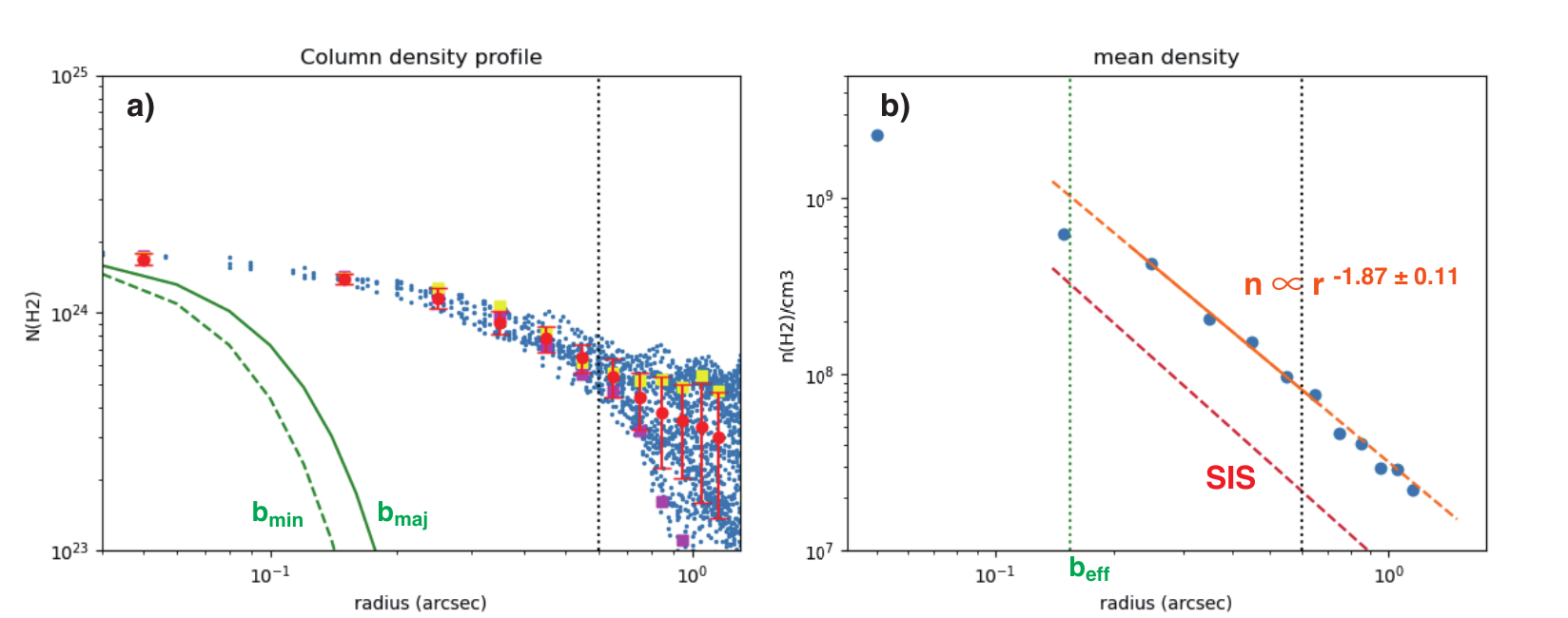}
\caption{a) Radial profile of the H$_2$ column density for the nucleus derived from the 1.1 mm continuum emission.
The red dots with error bars are the azimuthal average of $N$(H$_2$) at 0\farcs1 bins.
The yellow and magenta squares are the H$_2$ column density profiles along the position angles of 120$^{\circ}$ and 30$^{\circ}$, respectively.
The green solid and dashed curves denote the synthesized beam along the major and minor axes, respectively.
b) Radial number density profile of the nucleus assuming spherical symmetry.
The orange line shows the best-fit power law profile to the data points at radii larger than the effective beam size (dotted vertical line).
The red line shows the density profile for the SIS  ($r^{-2}$) with an effective sound speed of 0.21 km s$^{-1}$. 
\label{fig:radial_profile}}
\end{figure*}

Figure \ref{fig:radial_profile}a shows the H$_2$ column density of the nucleus as a function of the distance from the peak.
The column density was derived using the 1.1 mm continuum data which achieved the highest angular resolution of $\sim$0\farcs15, allowing the nucleus to be spatially resolved.
In order to minimize contamination due to the emission from the filament, the pixels in the eastern half of the nucleus (with the position angles from 30 $^{\circ}$ to 210 $^{\circ}$) were used for the analysis.
The dust temperature was assumed to be constant at $T_{\rm dust} = $11 K, which is derived from the hf fitting of the N$_2$D$^+$ observed at $\sim$1.2\arcsec resolution.
The dust mass opacity, $\kappa_{\rm 260 GHz}$, was adopted to be 0.0145 cm$^{2}$g$^{-1}$ for the gas density of 10$^8$ cm$^{-3}$ \citep{Oss94}.

Assuming that the nucleus is spherically symmetric and consists of concentric shells, the radial density profile  was derived from the column density profile.
First, the column densities are azimuthally averaged in each radial bin with a 0\farcs1 width (the red dots in Figure \ref{fig:radial_profile}a).
Then, the mass in each annulus is calculated from the averaged column density and the area of the annulus.
The mass enclosed within a radius $r$ is derived by summing up the masses of the annuli inside $r$.
The mean volume density in each shell is determined from the mass of annulus divided by the volume of the shell with the same radius and width of the corresponding annulus.
Figure \ref{fig:radial_profile}b shows the mean density as a function of radius.
The volume density increases from $\sim$10$^{8}$ cm$^{-3}$ at $r =$0\farcs6 (230 au) to $\sim$2$\times$10$^{9}$ cm$^{-3}$ at the center.
The density profile at $r >$ 0\farcs2 (80 au) is fit by a power law with an index of $-$1.87${\pm}$0.11 (orange solid line in Figure \ref{fig:radial_profile}b), which is close to $r^{-2.0}$.
It should be noted that the deviation from spherical symmetry is significant at $r >$ 0\farcs7 because of the filamentary distribution of the continuum emission.
The column density along the filament (yellow squares in Figure \ref{fig:radial_profile}a) does not decrease at $r >$ 0\farcs7 due to the low-level continuum emission that extends to the east of the nucleus, while the column density across the filament drops sharply (magenta squares in Figure \ref{fig:radial_profile}a).
Therefore, the data points at $r >$ 0\farcs7 were not used for the power-law fitting because the mean density values in the outer area are not reliably dominated by the nucleus. 
The innermost two data points locate below the {
\bf $r^{-1.87}$} line, implying the possible density flattening at the center.
However, due to the limited angular resolution of $\sim$0\farcs15, it is difficult to determine the density profile at $r <$ 0\farcs2 region.
Although the density profile of the nucleus, $r^{-1.87}$, is close to that of the Singular Isothermal Sphere (SIS) with ${\rho}(r) = ({\Delta}v_{\rm tot}^2/2 {\pi} G) r^{-2}$, where ${\Delta}v_{\rm tot}$ is an effective sound speed \citep{Shu77}, the density scaling is not consistent with that of a SIS.
Given the line width toward the nucleus obtained from the hf spectral fitting, which is 0.29 km s$^{-1}$, the total velocity dispersion derived by the equation (\ref{eq:dv_tot}) becomes 0.22 km s$^{-1}$.
The density profile of the SIS utilizing this ${\Delta}v_{\rm tot}$ is shown in the red dashed line in Figure \ref{fig:radial_profile}b.
It is evident that the density of the nucleus at each radius is higher than that of the SIS value by a factor of $\sim$3.7.
This implies that the gravity is stronger than thermal and turbulent pressure at every radii in the nucleus.

It should be noted that the assumed gas temperature, which is derived from the hf spectral fitting of N$_2$D$^+$ observed at lower resolution, does not always represent the gas temperature in the nucleus.
If the temperature in the nucleus decreases toward the small radius as in the case of L1544 \citep{Cra07}, the actual density profile could be steeper than the derived value.
If the nucleus is collapsing dynamically, the isothermal condition can be maintained by the continuous inflow of the gas from the outer region.
On the other hand, if the nucleus contains an internal heating source, the temperature could exceed 11 K.
In such a case, the density profile in the central region becomes flatter.

The virial parameter of the nucleus in the absence of magnetic support is calculated by 
\begin{equation}
    {\alpha} = \frac{5R {\Delta}v_{\rm tot}^2}{aGM},
    \label{eq:alpha}
\end{equation}
where R is a radius of the nucleus, ${\Delta}v_{\rm tot}$ is a total one-dimenstional velocity dispersion derived from equation (\ref{eq:dv_tot}), $a$ is a constant that depends on the density distribution, $G$ is the gravitational constant, and $M$ is the mass of the nucleus \citep{Ber92}.
The constant $a$ is given as $a = (1 - n/3) /(1 - 2n/5)$ for the case of the power-law density profile of ${\propto}$ $ r^{-n}$, and is 5/3 for  $n=$2.
The radius of the nucleus was assumed to be $r=$0\farcs6 (230 au; dashed open circle in Figure \ref{fig:cont}d), within which the contamination of the filament component is less significant.
The mass enclosed in a radius of 230 au (0\farcs6) is estimated to be $\sim$0.1 $M_{\odot}$.
Utilising the total velocity dispersion in the nucleus, which is 0.22 km s$^{-1}$, the virial parameter $\alpha$ is calculated to be 0.39, which implies that the nucleus cannot be supported by turbulence.

\citet{Kau13} argued that the cores with low virial parameters are not collapsing.
The velocity dispersions in the collapsing cores should increased due to the inward motions, resulting in the virial parameter $\alpha$ to be closer to 2a (i.e. 10/3 for n = 2).
The N$_2$D$^+$ line does not exhibit the signature of such line broadening toward the nucleus.
The absence of line broadening in N$_2$D$^+$ may be attributed to the inability of this line to trace the gas in the vicinity of the nucleus due to molecular depletion. 
As described in section \ref{subsec:N2Dp_abundance}, the N$_2$D$^+$ abundance decreases toward the nucleus, suggesting that this molecule disappears from the gas in the immediate vicinity of the nucleus. 

On the other hand, the nucleus could be supported against collapse by magnetic field.
The maximum mass, $M_{\Phi}$, supported by magnetic field can be calculated by 
\begin{equation}
    M_{\Phi} = c_{\Phi}\frac{{\pi}BR^2}{G^{1/2}},
    \label{eq:Mphi}
\end{equation}
where B is the magnetic field strength and c$_{\Phi}$ is 0.09 for a centrally condensed cloud \citep{Tom88}.
Using Equation (\ref{eq:Mphi}), the required field strength to support the nucleus with a mass of $\sim$0.1 $M_{\odot}$ and a radius of 230 au is estimated to be $\sim$15 mG.
Even though the contribution of the turbulence is taken into account, very strong field of $\sim$10 mG is required to support the nucleus.
The field strength measured for typical protostellar envelopes with a mean density of a few times 10$^7$ cm$^{-3}$ is on the order of a few mG \citep{Aso21,Gir06,Hul17}.
Under the flux-freezing condition, the scaling relation between the magnetic field strength and density is $B{\propto}{\rho}^{1/2}$ \citep{Cru12,Tri15,Hen19}. 
Therefore, the field strength can be $\gtrsim$10 mG in a region with a mean density of a few times 10$^8$ cm$^{-3}$ if the magnetic field is frozen in the matter.

\section{Discussion} \label{sec:discussion}

\subsection{Chemical stratification in G208-N2 
\label{subsec:stratification}}

As shown in \citet{Sah21}, the spatial distribution of the N$_2$D$^+$ emission shows good correlation with that of the continuum emission.
However, the column density of N$_2$D$^+$ exhibits significantly different distribution with respect to the H$_2$ column density due to the presence of the high opacity component in the northwestern part of the filament.
This implies that G208-N2 consists of two components; one is the continuum filament harboring the nucleus (filament component) and the other is the one with high N$_2$D$^+$ column density (N$_2$D$^+$ abundant component) located to the northwestern part.
Although two components are not clearly identified in the channel map of N$_2$D$^+$ (Figure \ref{fig:chmap}) due to the blending of the multiple hyperfine components, the radial velocity distribution (Figure \ref{fig:N2Dp_parameters}c) suggests that these two components have slightly different radial velocities, i.e. $V_{\rm LSR}{\sim}$10.8--10.9 km s$^{-1}$ in the filament component and $V_{\rm LSR}{\sim}$11.0--11.1 km s$^{-1}$ in the N$_2$D$^+$abundant component.
The $X(N_2D^+)$ value of the N$_2$D$^+$ abundant component is higher than $>$10$^{-10}$, which is comparable to the fractional abundance of N$_2$H$^+$, 3.0$\times$10$^{-10}$, in dense cores \citep{Cas02}.
The N$_2$D$^+$ abundance in the western part of the filament component is not clear because this part spatially overlaps with the N$_2$D$^+$ abundant component.
The N$_2$D$^+$ abundance drops significantly to  $<$2${\times}$10$^{-11}$ in the vicinity of the nucleus.
The possible origin of this abundance drop is discussed in the next subsection.

As described in section \ref{subsec:dense_gas}, the C$^{18}$O emission is completely missing in the dense gas of G208-N2.
Since the total power data are not combined, the missing C$^{18}$O flux should be significant.
However, the missing C$^{18}$O flux mainly comes from the spatially extended component along the line of sight, which is not associated with the dense gas in this core.
If the extended C$^{18}$O emission along the line of sight has a large optical depth, it could obscure the emission from the G208-N2 core.
In order to estimate the optical depth of the extended C$^{18}$O emission, we used the spectra obtained with the Total power (TP) array having a beam size of $\sim$30\arcsec.
As shown in Figure \ref{fig:TPspectra}, the brightness temperature of the C$^{18}$O spectrum observed toward the nucleus position is $\sim$8.8 K, while that of the $^{12}$CO spectrum is $\sim$26 K.
Assuming that the excitation temperature of the extended C$^{18}$O emission{ , $T_{\rm ex}$(C$^{18}$O$_{\rm fg}$),} is the same as the peak brightness temperature of the $^{12}$CO emission, the optical depth of the extended C$^{18}$O emission{ , $\tau$(C$^{18}$O$_{\rm fg}$),} is estimated to be $\sim$0.55.
{ If it is the case, the foreground C$^{18}$O emission does not completely obscures the emission from the G208-N2 core.
On the other hand, $\tau$(C$^{18}$O$_{\rm fg}$) becomes larger than unity for $T_{\rm ex}$(C$^{18}$O$_{\rm fg}$)~$<$ 19 K.
if $T_{\rm ex}$(C$^{18}$O$_{\rm fg}$) is comparable to that of the dense gas, i.e. $\sim$ 15 K, the foreground cloud with  $\tau$(C$^{18}$O$_{\rm fg}$)$\sim$2  obscures $\sim$90\% of the emission from the dense core behind it.
This scenario can explain the absence of C$^{18}$O emission from the G208-N2 core.
Nevertheless, such an extended foreground component with large optical depth might obscure the C$^{18}$O emission from the neighbouring protostellar cores, MMS3 and MMS5, located at  $\sim$33\arcsec and $\sim$26\arcsec distances  from the nucleus.
Given the C$^{18}$O flux densities observed in MMS3 and MMS5 that correspond to the brightness temperatures of $\sim$10 K and $\sim$15 K, respectively \citep{Mor21,Mat19}, it is unlikely that a substantial portion of the emission from these cores is obscured. }

{It is more plausible that the lack of C$^{18}$O emission within the dense gas of G208-N2 is attributable to substantial CO depletion.}
The 3 $\sigma$ upper limit of the C$^{18}$O intensity is estimated to be $\sim$90 mJy beam$^{-1}$, which corresponds to $\sim$1.58 K in the brightness temperature unit.
Assuming that the C$^{18}$O emission has the same line width as N$_2$D$^+$, i.e. $\sim$0.3 km s$^{-1}$, the upper limit of the integrated intensity becomes 0.48 K km s$^{-1}$.
Under the LTE condition with an excitation temperature of 11 K, and the optically thin line emission, the column density of C$^{18}$O is calculated to be $N$(C$^{18}$O) $<$ 2.8 $\times$10$^{14}$ cm$^{-2}$.
Since the H$_2$ column density derived from the 1.3 mm continuum emission is higher than 2 $\times$10$^{23}$ cm$^{-2}$ in the entire region of G208-N2, the upper limit of the C$^{18}$O abundance, $X$(C$^{18}$O), is estimated to be $\sim$10$^{-9}$.
The derived C$^{18}$O abundance is more than two orders of magnitude lower than the C$^{18}$O abundance value often used in the interstellar medium, (1.7--2)$\times$10$^{-7}$ \citep{Wan80,Fre82}.
This implies that the C$^{18}$O (CO) depletion factor in G208-N2 is greater than 100 in the entire region, which is one order of magnitude larger than those of the nearby prestellar cores in the same size scale \citep{Cra05}.
The C$^{18}$O (CO) depletion factor toward the nucleus is even higher, and is $>$ 350.

The DCO$^+$ emission is enhanced at the southern edge of the triangle-shaped dense gas.
Since DCO$^+$ formation requires CO in the gas phase through the reactions of H$_2$D$^+$ $+$ CO $\rightarrow$ DCO$^+$ $+$ H$_2$ and N$_2$D$^+$ $+$ CO $\rightarrow$ DCO$^+$ $+$ N$_2$,
the enhanced DCO$^+$ emission implies that the southern edge of the dense gas is heated above the sublimation temperature of CO , $\sim$ 25 K.
The faint C$^{18}$O emission is seen in the velocity range from $V_{\rm LSR}$ = 11.3--11.8 km s$^{-1}$ in the channel map (Figure \ref{fig:N2Dp_C18O}).
The location of this component is consistent with that of the DCO$^+$ emission shown in Figure \ref{fig:lines}c.

The oval-shaped morphology of the H$_2$CO emission suggests that H$_2$CO
traces the external layer surrounding the continuum filament.
H$_2$CO can be formed through both grain-surface chemistry and gas-phase chemistry.
In the gas-phase route,  H$_2$CO is formed through reactions of hydrocarbons such as CH$_2$ and CH$_3$ with oxygen atoms \citep{Yam17}.
In the case of the G208-N2, this route is unlikely because carbon is completely locked to CO and frozen onto grain surface in such a dense and cold environment.
In the region where CO is frozen out, H$_2$CO is formed on the grain surfaces by successive hydrogenation of CO.
The H$_2$CO in the ice can be released in the gas phase when the grains are heated to its sublimation temperature. 
However, the sublimation temperature of H$_2$CO is $\sim$40 K for the desorption energy of 2050 K \citep{Gar06}, which is much higher than the gas temperature in G208-N2.
UV induced photodesorption is also unlikely, because UV radiation cannot reach the surface of the continuum filament.
Possible desorption mechanisms in the external layer could be grain heating by means of the impact of cosmic rays \citep{Her06} or the formation energy released by the exothermic reaction \citep{Gar06b}.

\subsection{Local gas heating in the vicinity of the nucleus}
\label{subsec:heating}

\begin{figure}
\epsscale{0.8}
\plotone{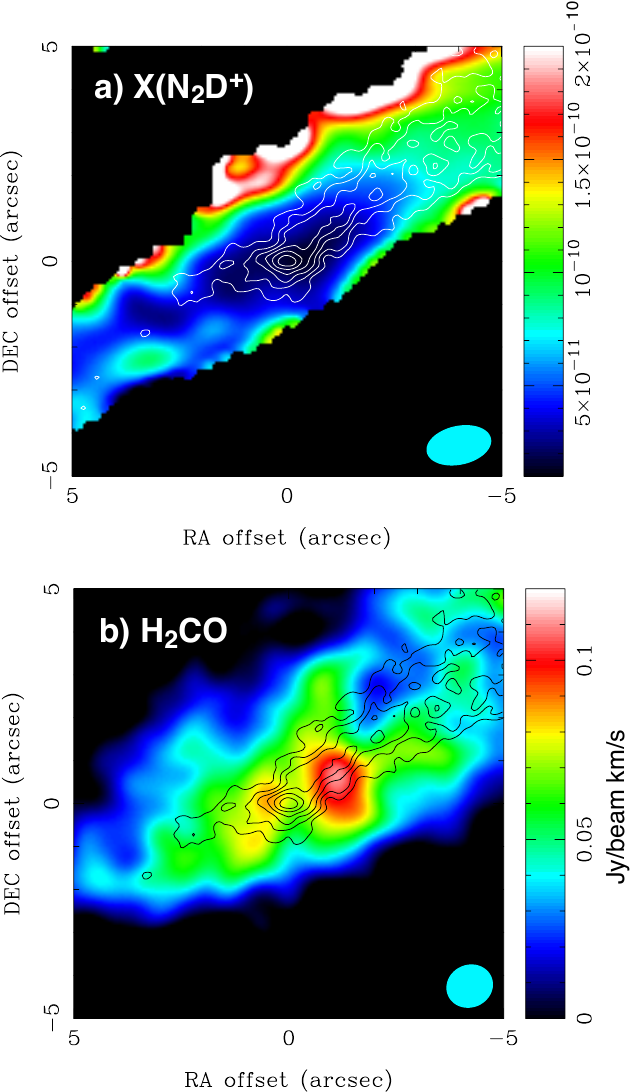}
\caption{a) Fractional abundance of N$_2$D$^+$ and b) integrated intensity of H$_2$CO 3$_{0,3}$--2$_{0,2}$ in the vicinity of the nucleus overlaid on the 1.3 mm continuum images drawn contours.
The contours are drawn every 3 $\sigma$ with the lowest contour level of 6 $\sigma$.
The 1 $\sigma$ level is 0.084 mJy beam$^{-1}$.
 The beam size of the 1.3mm continuum is 0\farcs38$\times$0\farcs30, while those of the N$_2$D$^+$ and H$_2$CO shown in the bottom-right of each panel are 1\farcs53$\times$0\farcs91 and 1\farcs10$\times$1\farcs03, respectively.
\label{fig:X_H2CO}}
\end{figure}

As shown in Figure \ref{fig:column_density}c and Figure \ref{fig:X_H2CO}a, the N$_2$D$^+$ abundance that is $\sim$8$\times$10$^{-11}$ in the western part of the filament decreases by a factor of $\sim$5 in the vicinity of the nucleus.
Since the density of the gas in the vicinity of the nucleus is extremely high, $>$10$^{7}$ cm$^{-3}$, N$_2$, a parent molecule of N$_2$D$^+$, is expected to freeze-out if the temperature is low.
The binding energy of N$_2$ measured in the laboratory experiments is $\sim$0.9 times of that of CO \citep{Obe05,Fay16}.
As a result, the N$_2$ freeze-out temperature is a few degrees lower than that of CO.
Although the clear anticorrelation between C$^{18}$O and N$_2$D$^+$ observed in G208-N2 implies that the N$_2$ freeze-out timescale is longer than that of CO, N$_2$ could also freeze-out in the region of the highest density.
In the center of the well-studied prestellar core, L1544, the temperature of which decreases toward the center \citep{Cra07}, the depletion of N$_2$D$^+$ is observationally confirmed \citep{Red19}.
Therefore, if there is no heating source embedded in the nucleus, the temperature drop and N$_2$ depletion could also happen in the nucleus of G208-N2.

Alternatively, the N$_2$D$^+$ abundance decreases in the vicinity of the nucleus because of the increased temperature due to the embedded heating source.
Although the nucleus is not detected in the wavebands shorter than 70$\mu$m (Fig. \ref{fig:PACS_SCUBA}), this does not exclude the presence of embedded heating source.
One of the FHSC candidates B1-bN is also not detected in these wavebands \citep{Pez12}.
Once the central source is formed, small amount of CO starts to sublimate, destroy N$_2$D$^+$, and decrease the N$_2$D$^+$ abundance.
Although the C$^{18}$O emission is not detected toward the nucleus, this does not completely rule out the presence of small amount of CO in the gas phase.
The upper limit of the CO abundance toward the nucleus derived from that of the C$^{18}$O abundance is $<$3$\times$10$^{-7}$, assuming the O$^{16}$/O$^{18}$ of 560 \citep{Wil94}.
This is much larger than the N$_2$D$^+$ abundance, 1.6$\times$10$^{-11}$ at the nucleus and $\sim$10$^{-10}$ in the region with high N$_2$D$^+$ abundance.
This scenario also explains the presence of the two H$_2$CO emission spots in the vicinity of the nucleus.
As shown in Figure \ref{fig:X_H2CO}b, two H$_2$CO emission spots, one at 1\farcs25 (490 au) northwest and the other at 0\farcs5 (200 au) east of the nucleus, are located in the region with low N$_2$D$^+$ abundance.
These H$_2$CO emission spots could be the signposts of the warm regions.

In the case of the thermal desorption of H$_2$CO due to the radiation from the central object, the H$_2$CO emission is expected to appear toward the nucleus.
Since the nucleus does not have a counterpart in the {\it Herschel} 70 $\mu$m image, the luminosity of the putative central object should be very low.
The upper limit flux of the nucleus at 70 $\mu$m derived from the rms noise level is $\sim$60 mJy.
The upper limit of the internal luminosity of the nucleus can be roughly estimated using the empirical relation between $L_{\rm int}$ and 70$\mu$m flux \citep{Dun08}.
Using this method, the upper limit of the internal luminosity of the nucleus is estimated to be 0.03 $L_{\odot}$.
The sublimation radius of H$_2$CO is expected to be $\sim$20--30 au for such a very low luminosity object \citep{Aik08}.
However, neither of the H$_2$CO peaks coincides with the position of the nucleus.
If the envelope surrounding the nucleus has cavities as in the cases of the envelopes in Class 0 protostars, the radiation from the central object can rather freely escape through the cavity.
Even so, it is very difficult for the very low luminosity source to heat the large area of 200--500 au to the sublimation temperature of H$_2$CO.

Instead of the thermal desorption, the H$_2$CO could be sublimated by means of slow shocks.
The two H$_2$CO peaks offset from the nucleus could be the outflow lobes as in the cases of the outflows from the youngest protostars, B1-bN and B1-bS \citep{Ger15}.
Since the outflows traced by the H$_2$CO emission are $\sim$5\arcsec ($\sim$1500 au) in B1-bN and $\sim$10\arcsec ($\sim$3000 au) in B1-bS, the 200--500 au offsets of the H$_2$CO peaks can be explained if they are tracing the compact outflow lobes from the central object.
The caveat to this scenario is the absence of the CO counterpart.
However, this could be due to the very low velocity of this putative outflow.
Since the H$_2$CO emission is detected only in the single channel with a velocity resolution of 1.55 km s$^{-1}$, the radial velocity of the putative outflow is $<$ 0.8 km s$^{-1}$.
Such a low velocity ``outflow'' could be completely overwhelmed by the CO emission from the cloud, which is resolved out by the interferometer.
Another possible scenario is the slow shocks resultant from the inflow.
The velocity gradient along the major axis (Figure \ref{fig:N2Dp_PV}a) reveals a gradual reduction in LSR velocities of the gas on either side of the nucleus.
Such a V-shaped configuration centered on the nucleus, can be attributed to the inflow of gas towards the nucleus, with the filamentary structure potentially undergoing a kink and inclination shift at the same location.
The H$_2$CO emission spot is brighter in the western side, as the shock is stronger due to the inflow of higher-density gas.
In order to study the nature of the H$_2$CO emission spots, high spectral resolution observations of H$_2$CO are necessary.

\subsection{Evolutionary stage of the nucleus}
\label{subsec:nucleus}

The nucleus in G208-N2 has a radius of $\sim$230 au and a mass of $\sim$0.1 $M_{\odot}$.
The nucleus has a very high central density of $\sim$3$\times$10$^9$ cm$^{-3}$.
Although the radial density profile,  $r^{-1.87}$, is close to $r^{-2}$ for the SIS, the density at each radius is higher than that of the SIS by a factor of $\sim$3.7.
The stability analysis implies that the nucleus is gravitationally unstable unless it is supported by strong magnetic field of $>$ 10 mG.
 The central density of the nucleus in G208-N2, i.e. $\sim$10$^9$ cm$^{-3}$, is comparable to that of SM1N in  Ophiuchus A \citep{Fri18}, although the nucleus in G208-N2 is three times more massive than SM1N.
In the case of Oph A SM1N, the  flatter density profile of $r^{-1.3}$ and the  detection of the blue- and redshifted CO emission with a velocity extent of a few km s$^{-1}$ imply the presence of the central embedded object, which is FHSC or early Class 0 stage \citep{Fri18}.
If the two emission spots seen in the H$_2$CO image are indeed indicators of the outflow, the nucleus may harbor its central source.
The upper limit of its internal luminosity derived from the upper limit 70 $\mu$m flux is 0.03 $L_{\odot}$, which is
in the range of the FHSC predicted from the theoretical models, $L$ ${\la}$ 0.25--0.1 $L_{\odot}$ \citep[e.g.][]{Omu07,Sai08,Com12,Vay12}.
The mass of the nucleus, $\sim$0.1 $M_{\odot}$, is not large enough to form a low-mass star unless the entire gas in the nucleus is converted into a star.
This implies the possibility that the central putative source could evolve into a brown dwarf.
However, the small velocity gradients along the filament from the both sides of the nucleus suggest that the nucleus can acquire additional gas from the filament and form a low-mas star.
Conversely, if the H$_2$CO emission spots result from inflow shocks, the nucleus has not yet formed its central object.
In this case, the nucleus can be in the prestellar stage on the verge of FHSC formation.

\subsection{Shock compression}
\label{subsec:compression}

 The P-V diagram of N$_2$D$^+$ across the filament (Figure \ref{fig:N2Dp_PV}b) exhibits a V-shaped pattern with broad redshifted wing emission at the southern edge where the line intensity drops sharply.
This implies that the dense gas in G208-N2 is compressed by the external pressure from the south.
The enhancement of the DCO$^+$ emission at the southern boundary also  supports the presence of shocks that heat the gas above the sublimation temperature of CO.
 Furthermore, the continuum filament, characterized by an average density of 10$^7$ cm$^{-3}$, exhibits a parallel alignment with the southern periphery of the dense gas traced by the N$_2$D$^+$ line. 
This also suggests that external pressure from the south compressed the dense gas within  G208-N2, leading to the formation of a denser filament within.
However, the origin of this external pressure is unclear. 
 The northern part of OMC-3, in which G208-N2 is embedded, resides along the southwestern periphery of the expanding bubble delineated by the [CII] emission driven by NGC 1977 \citep{Pab20}, suggesting that this filamentary cloud was formed under the influence of the pressure from the bubble.
Nevertheless, the pressure exerted by this bubble originates from the northeastern direction, contrasting with the pressure coming from the south.
It is also unlikely that the pressure localized to G208-N2 originates from such a gigantic bubble with a radius of 1.6 pc.
There is a flat spectrum YSO HOPS 89 \citep{Meg12,Fur16} at $\sim$17'' southwest of the nucleus.
However, this YSO does not show any signature of outflow that can account for the external pressure.
The Class 0 protostar HOPS 88 \citep{Meg12,Fur16} embedded in the neighbouring core MMS5 is driving a bipolar outflow along the east-west direction \citep{Aso00,Wil03,Tak08}.
Although the western lobe of this outflow overlaps with the location of G208-N2, this lobe is blueshifted with respect to the cloud systemic velocity. 
Therefore, this outflow is unlikely to be the origin of the external pressure that produces the redshifted N$_2$D$^+$ wing emission.
The moment 1 map of $^{13}$CO 1--0 \citep{Kon18} reveals that the dense gas in OMC-3 having a radial velocity of $\sim$10.5--11.0 km s$^{-1}$ is embedded in the diffuse gas with $\sim$12 km s$^{-1}$.
Therefore, the origin of the external pressure could be diffuse molecular gas surrounding OMC-3 region.

\subsection{End-dominated collapse in the filamentary structure \label{end-dominated collapse}}

In the case of the filaments with finite lengths, in which gas motion along the filament becomes important, their collapse is different from that of the filament with infinite length.
It is known that collapsing finite-sized filaments form dense condensations at the filament edges by means of gravitational focusing \citep[e.g.][]{Bas83,Bur04,Pon12,Cla15}.
 On the basis of analytical calculations of the collapse timescales, \citet{Pon12} revealed that this end-dominated collapse mode becomes more important with increasing aspect ratio.
The dense condensations formed at the ends of the filament can collapse locally before the entire filament converges toward its center when the local collapse timescale is significantly shorter than the global collapse timescale \citep{Pon11}.

 This end-dominated collapse scenario is applicable to both the OMC-3 ridge, in which G208-N2 is embedded, and the continuum filament in G208-N2, in which the nucleus is embedded.
 The northern part of the OMC-3 ridge from MMS1 to MMS6 with a length of $\sim$0.35 pc and a width of $\sim$0.06 pc has an aspect ratio of $\sim$6.
According to \citet{Pon12}, the timescale of the global collapse with an end-dominated mode is $\sqrt{32A/{\pi}^2}\,{\tau}_{ff}$, where $A$ is an aspect ratio and ${\tau}_{ff}$ is a free-fall timescale.
The free-fall timesceale of this ridge with a mean density of $\sim$4$\times$10$^5$ cm$^{-3}$ \citep{Sch21} is estimated to be $\sim$4.9$\times$10$^4$ yr.
Hence the collapse timescale for the OMC-3 ridge with an aspect ratio of $\sim$6 becomes $\sim$2.2$\times$10$^5$ yr.
The timescale for the global collapse could be longer because the magnetic field perpendicular to the OMC-3 ridge \citep{Mat01,Hou04,Li22b} provides support along the major axis of the ridge.
Once the dense condensation with $>$10$^6$ cm$^{-3}$ is formed at the edge, the free-fall timescale of this condensation, $\sim$3.3$\times$10$^4$ yr is much shorter than the global collapse timescale.
Therefore, the condensation formed at the end can collapse locally.

 On the basis of one-dimenstional hydrodynamic simulations, \citet{Han94} showed that the formation of a dense condensation can propagate along the filament; if the local collapse mode works, the condensation formed at the edge could be isolated from the parent filament.
Consequently, the parent filament could have a new edge where a subsequent condensation takes form.
If the formation of condensation occurs at both ends of the filament, it propagates inward at approximately the effective sound speed.
\citet{Tak13} argued that this model can explain the evolutionary scenario of the OMC-3 ridge, which harbors the youngest condensation, G208-N2, in the middle of the chain of condensations.
They estimated the timescale required for the fragmentation to propagate from the two edges of the ridge to G208-N2 at the center to be $\sim$3$\times$10$^5$ yr.

 Within the G208-N2  core, the compact nucleus is located at the eastern end of the continuum filament.
Although the filamentary cloud itself extends to the southeast of the nucleus, the H$_2$ column density  exhibits a significant decrease of $\sim$50\% at $\sim$1\arcsec (400 au) east of the nucleus.
Such a configuration can  also be explained in the context of the end-dominated collapse of the finite-sized filament.
 \citet{Han94} and \citet{Pon11} examined the formation of condensations at the end of the filament with a gradual density decrease.
They found that the end-collapse occurs in the filaments with tapered edges, although the collapse timescale increases as the length of the tapered region increases.
The filament in G208-N2 contains only one condensation at its eastern edge instead of two condensations at both edges.
 The absence of condensation in the western edge of the G208-N2 filament can be explained by the gradual density decrease in the northwest.
The numerical simulation of \citet{Bur04}  also shows that the surface density gradient leads to the formation of a single condensation at the dense end.

 The timescale of the end-dominated global collapse of the continuum filament  with a mean density of $\sim$5.9$\times$10$^7$ cm$^{-3}$ and an aspect ratio of $\sim$8 is estimated to be $\sim$2.2$\times$10$^4$ yr.
On the other hand, the free-fall timescale of the nucleus with a mean density of 3.1--8.5$\times$10$^8$ cm$^{-3}$ is 1200--2000 yr, which is one order of magnitude shorter than that of the continuum filament.
Hence, the nucleus can collapse locally and form a protostar within.

 The quiescent and extremely dense object, akin to the nucleus in G208-N2 is rarely observed in nearby star-forming regions owing to its extremely short free-fall timescale.
The presence of such a short-lived object within G208-N2 can be attributed to the filamentary configuration of G208-N2 itself, nestled within the larger-scale filamentary structure of the OMC-3 ridge.
Due to this filament-in-filament configuration, the collapse timescale of G208-N2 surpasses that of neighboring cores within the OMC-3 ridge.
As previously discussed in section \ref{subsec: density profile}, the magnetic field strength may prove sufficient to counteract gravitational collapse if the nucleus was formed under the flux-freezing condition.

\section{Summary}
\label{sec:summary}

\begin{figure}
\epsscale{1.2}
\plotone{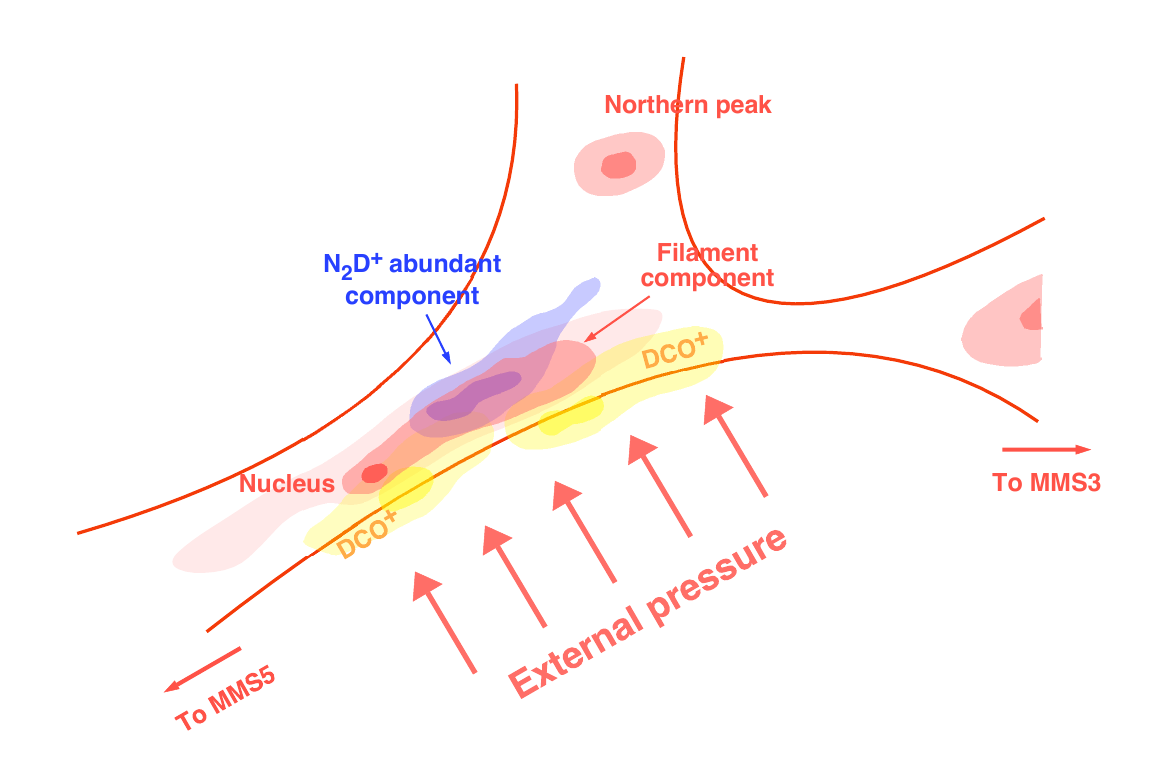}
\caption{Schematic diagram of G208-N2.
 The line-of-sight is perpendicular to the plane.
\label{fig:sketch}}
\end{figure}

We have studied the internal structure of the prestellar core G2008-N2 in the OMC-3 region using the ALMA in the 1.3 mm and 1.1 mm wavebands.
 The high resolution ALMA images have unveiled an extraordinarily dense nucleus nestled within this filamentary-shaped prestellar core. 
The main results are summarized as the following:
\begin{enumerate}
    \item The continuum emission from G208-N2 shows a narrow filamentary structure having a size of 4700 au $\times$ 590 au with an aspect ratio of $\sim$8.
    The position angle of this continuum filament is $\sim$120$^{\circ}$, which is roughly same as that of the OMC-3 ridge.
    The H$_2$ column density exceeds 5$\times$10$^{23}$ cm$^{-2}$ in the area of the continuum filament.
    The mass and mean density of this filament is $\sim$ 0.59 $M_{\odot}$ and  5.9 $\times$ 10$^7$ cm$^{-3}$, respectively.
    The line mass of the continuum filament,  1.67$\times$10$^{16}$ g cm$^{-1}$ (= 25.9 $M_{\odot}$), is  slightly larger than  the critical line mass of the isothermal cylinder at 11 K, suggesting that the filament is close to the critical state between stable and unstable.
    \item The continuum filament  harbors a compact ``nucleus'' with a radius of $\sim$150--200 au and a mass of $\sim$0.1 $M_{\odot}$ at its eastern end.
    The volume density at the center of the nucleus is $\sim$2 $\times$ 10$^9$ cm$^{-3}$.
   Such a compact and exceedingly dense object has been discovered within the prestellar core for the first time.
    There is no sign of CO outflow localized to this nucleus.
    \item 
    The nucleus has a radial density profile of  $r^{-1.87{\pm}0.11}$. 
    Although this density profile is close to $r^{-2}$ for the SIS, the density scaling is higher than that of the SIS value by a factor of $\sim$3.7.
    The virial parameter of the nucleus, determined from the velocity dispersion of the N$_2$D$^+$ 3--2 line, is remarkably low at 0.39.
    These imply that the gravity is dominant over the pressure everywhere in the nucleus unless the nucleus is strongly supported by magnetic field exceeding 10 mG.
    \item The molecular line data exhibit significant chemical stratification in G208-N2.
    The complete absence of C$^{18}$O emission implies CO depletion in the dense gas of G208-N2.
    The CO depletion factor is greater than 100 in the entire region, and $>$ 350 toward the nucleus.
    Although the spatial distribution of the N$_2$D$^+$ emission shows good correlation with that of the continuum emission,  the column density of N$_2$D$^+$ is enhanced on the north of the continuum filament.
    This implies that G208-N2 consists of two components; one is the continuum filament with high H$_2$ column density and the other is the one with high N$_2$D$^+$ column density.
    The DCO$^+$ emission is enhanced at the southern edge of the continuum filament, implying the external heating from the south.
    A schematic diagram of G208-N2 is given in Figure \ref{fig:sketch}.
    \item The redshifted N$_2$D$^+$ wing emission and the enhancement of the DCO$^+$ emission at the southern boundary of the triangle-shaped dense gas in suggest that the dense gas is compressed and heated by the shock from the south, although the origin of the external pressure is not identified.    
    \item The N$_2$D$^+$ emission does not peak toward the nucleus.
    As a results,
    the N$_2$D$^+$ abundance exhibits the local minimum toward the nucleus.
    This could imply either the depletion of N$_2$, a parent molecule of N$_2$D$^+$, due to low temperature and high density 
    or a limited release of CO from the dust, potentially attributed to the presence of an internal heating source.
    \item There are two H$_2$CO emission spots at $\sim$200 au east and $\sim$500 au west of the nucleus, which could be the locations of slow shocks caused by either low-velocity outflow or inflow.
    If the two emission spots are the signature of low-velocity outflow, the nucleus may harbor its central source, which could be a candidate for the FHSC.
    Conversely, if the emission spots results from inflow shocks, the nucleus is likely to be in the prestellar stage on the verge of FHSC formation.
   
    \item The nucleus is formed at the end of the continuum filament  that is embedded in the middle of the larger scale filamentary structure in the northern part of OMC-3.
    This implies that the gravitational focusing in the finite-sized filament plays an important role in  the evolution of both the OMC-3 ridge and the continuum filament within G208-N2.
    The presence of the extremely dense nucleus, which is a short-lived object, can be attributed to the filament-in-filament configuration, leading to an evolutionary timescale longer than that of neighboring cores within the OMC-3 ridge.
       
\end{enumerate}

\acknowledgments

This paper makes use of the following ALMA data:ADS/JAO.ALMA\#2015.1.00341.S and \#2018.1.00302.S.
ALMA is a partnership of ESO (representing its member states), NSF (USA) and NINS (Japan), together with NRC (Canada), NSC and ASIAA (Taiwan), and KASI (Republic of Korea), in cooperation with the Republic of Chile.
The Joint ALMA Observatory is operated by ESO, AUI/NRAO, and NAOJ.
We thank Dr. A. Trejo for helpful suggestions in data calibration.
N.H. acknowledge support from the National Science and Technology Council (NSTC) with grants NSTC 110-2112-M-001-048 and NSTC 111-2112-M-001-060.
D.S. acknowledges the support from Ramanujan Fellowship (SERB) and PRL, India.
K.T. was supported by JSPS KAKENHI (Grant Number JP20H05645).
This work has been supported by the National Key R\&D Program of China (No. 2022YFA1603100). 
T. L. acknowledges the supports by National Natural Science Foundation of China (NSFC) through grants No.12122307 and No.12073061, the international partnership program of Chinese Academy of Sciences through grant No.114231KYSB20200009, and Shanghai Pujiang Program 20PJ1415500.
D.J.\ is supported by NRC Canada and by an NSERC Discovery Grant.
L.B. gratefully acknowledges support by ANID BASAL project FB210003.
H.-L. Liu is supported by Yunnan Fundamental Research Project (grant No.\,202301AT070118).
The work of MGR is supported by NOIRLab, which is managed by the Association of Universities for Research in Astronomy (AURA) under a cooperative agreement with the National Science Foundation.

%

\vspace{5mm}
\facilities{ALMA}


\software{astropy \citep{Ast13},  
          CASA \citep{McM07}, 
          Miriad \citep{Sau95}
          }



\newpage
\clearpage

\appendix
\restartappendixnumbering
\section{Channel maps}
\label{sec:app_chmap}

\begin{figure*}
\epsscale{0.9}
\plotone{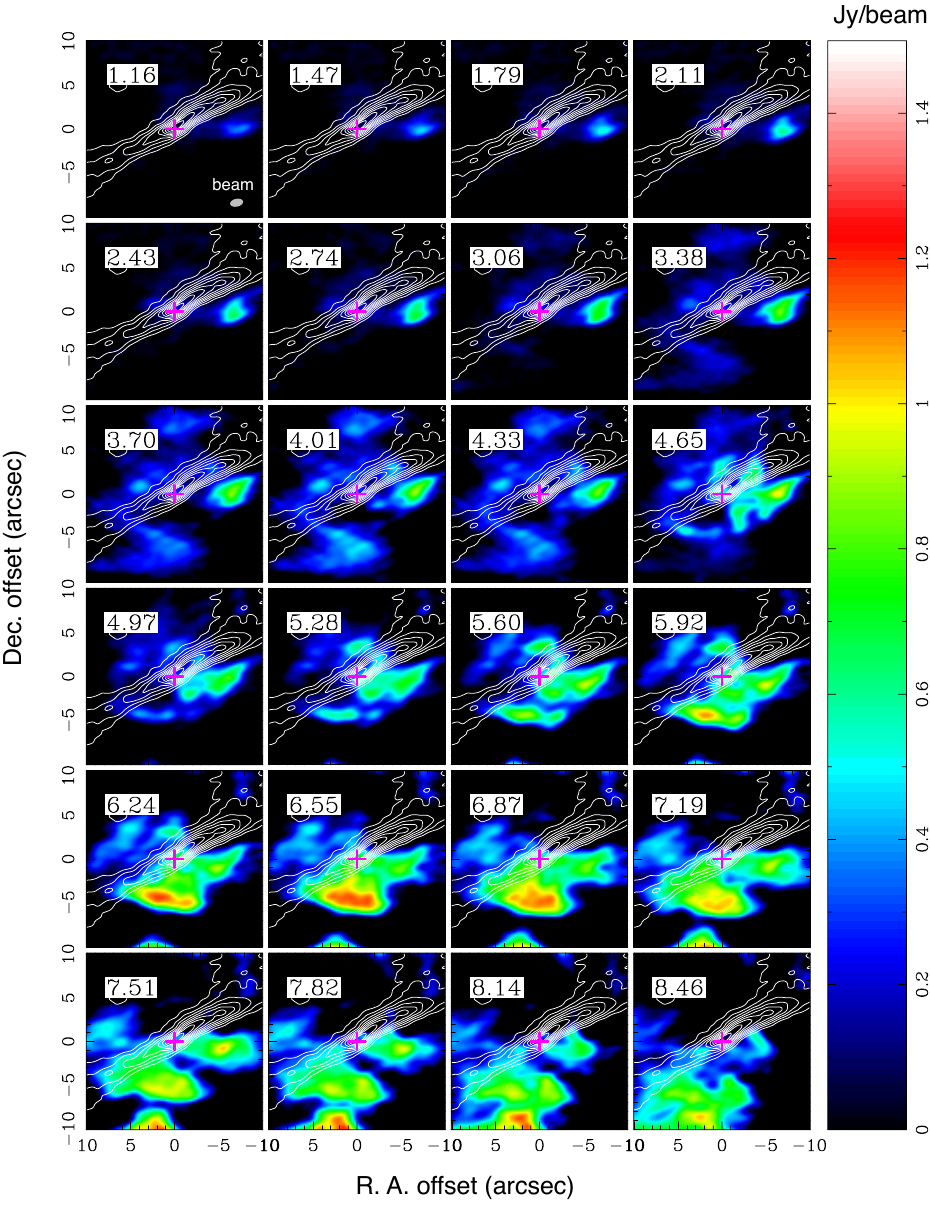}
\caption{Channel map of the CO 2--1 emission line at 0.32 km s$^{-1}$ resolution in color overlaid on the 1.3 mm continuum image in contours.
The velocity range is from 1.16 to 8.46 km s$^{-1}$.
Contours are drawn every 10\% level of the pdak flux density.
The magenta cross denotes the position of the nucleus.
\label{fig:CO_chmap1}}
\end{figure*}

\begin{figure*}
\epsscale{0.9}
\plotone{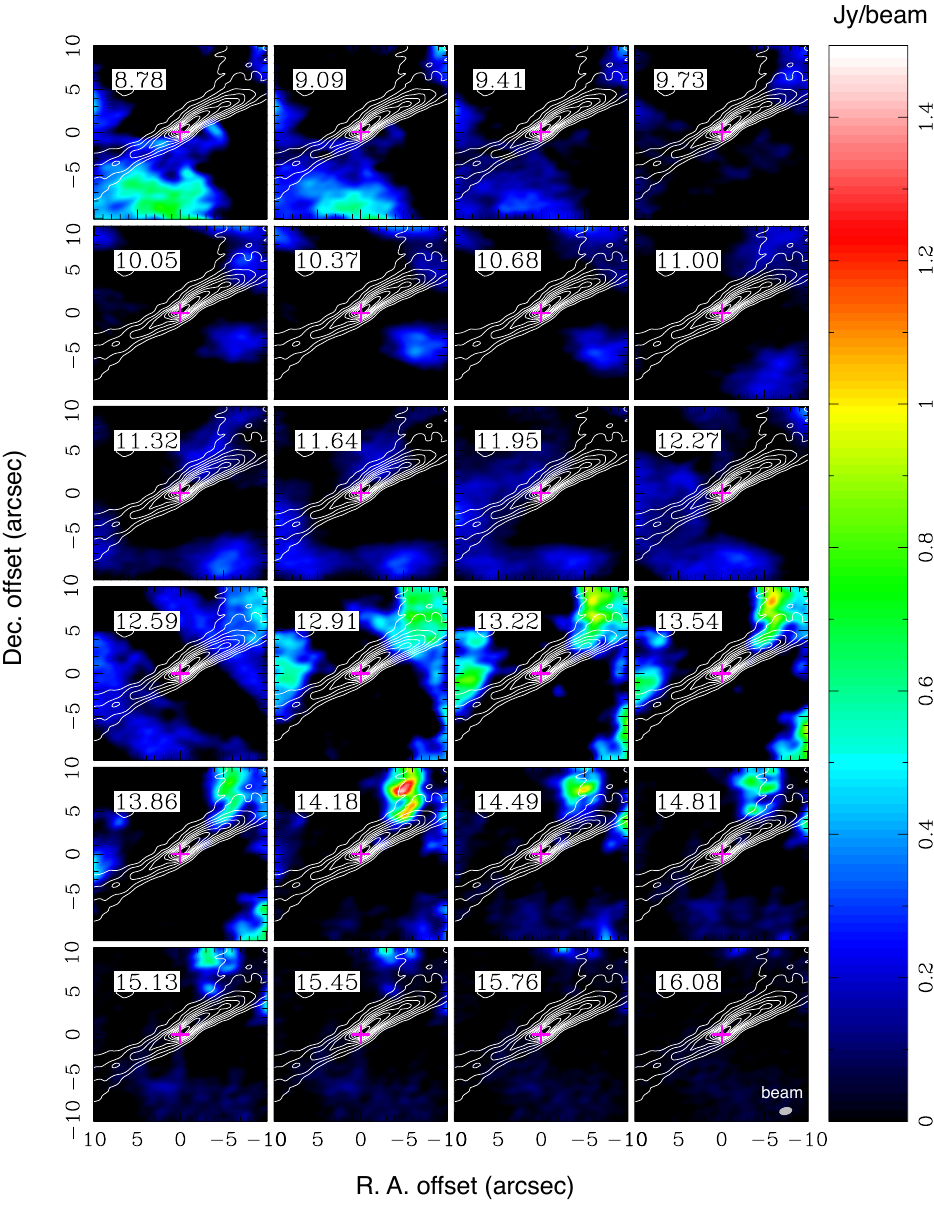}
\caption{Channel map of the CO 2--1 from $V_{\rm LSR}$ = 8.78 to 16.08 km s$^{-1}$.
\label{fig:CO_chmap2}}
\end{figure*}

\begin{figure*}
\plotone{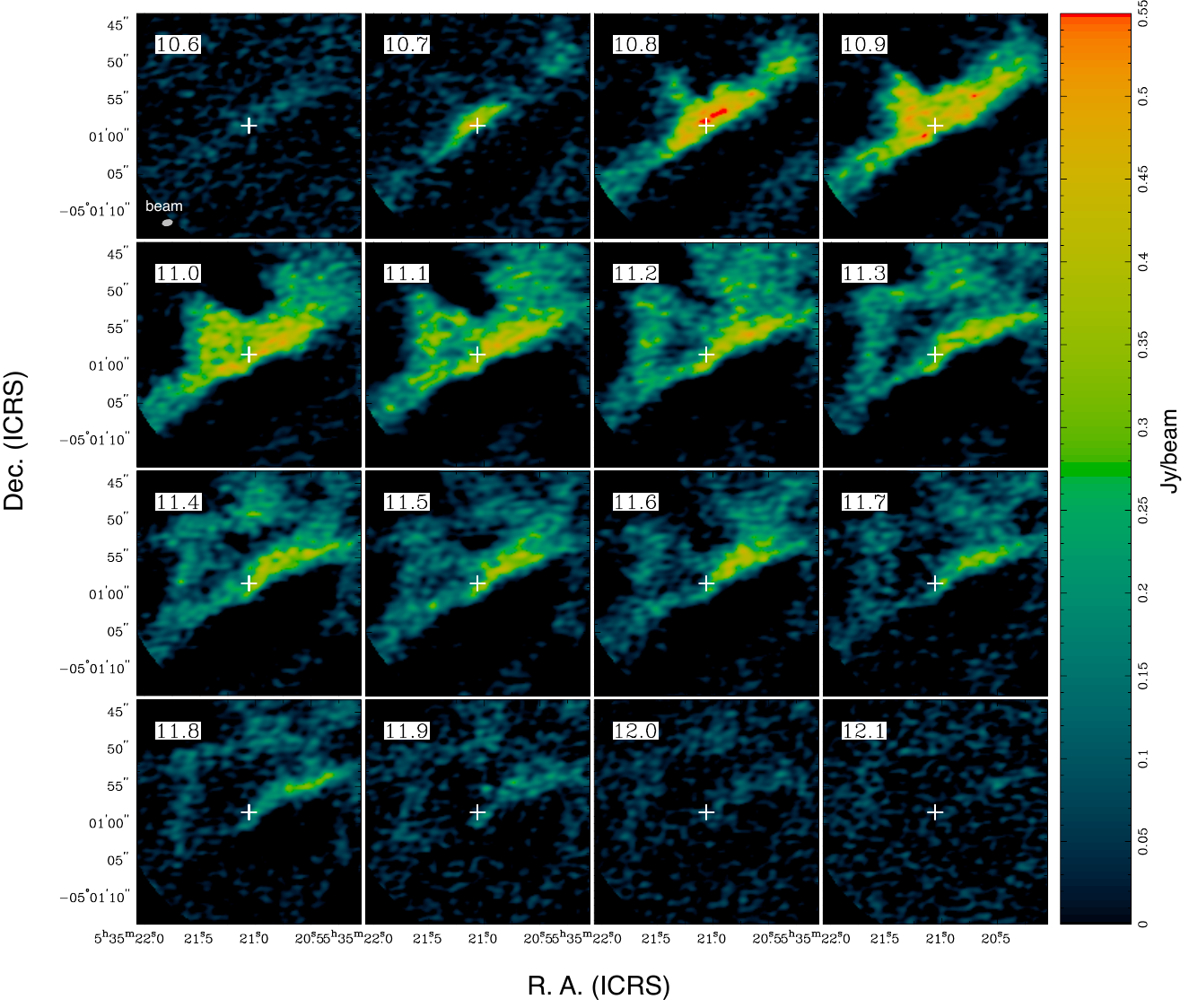}
\caption{Channel map of the N$_2$D$^+$ 3--2 emission line.
The velocity resolution in 0.1 km s$^{-1}$.
The LSR velocity refers to the rest frequency of 231.32187 GHz.
The white cross denotes the position of the nucleus.
\label{fig:chmap}}
\end{figure*}

\begin{figure*}
\plotone{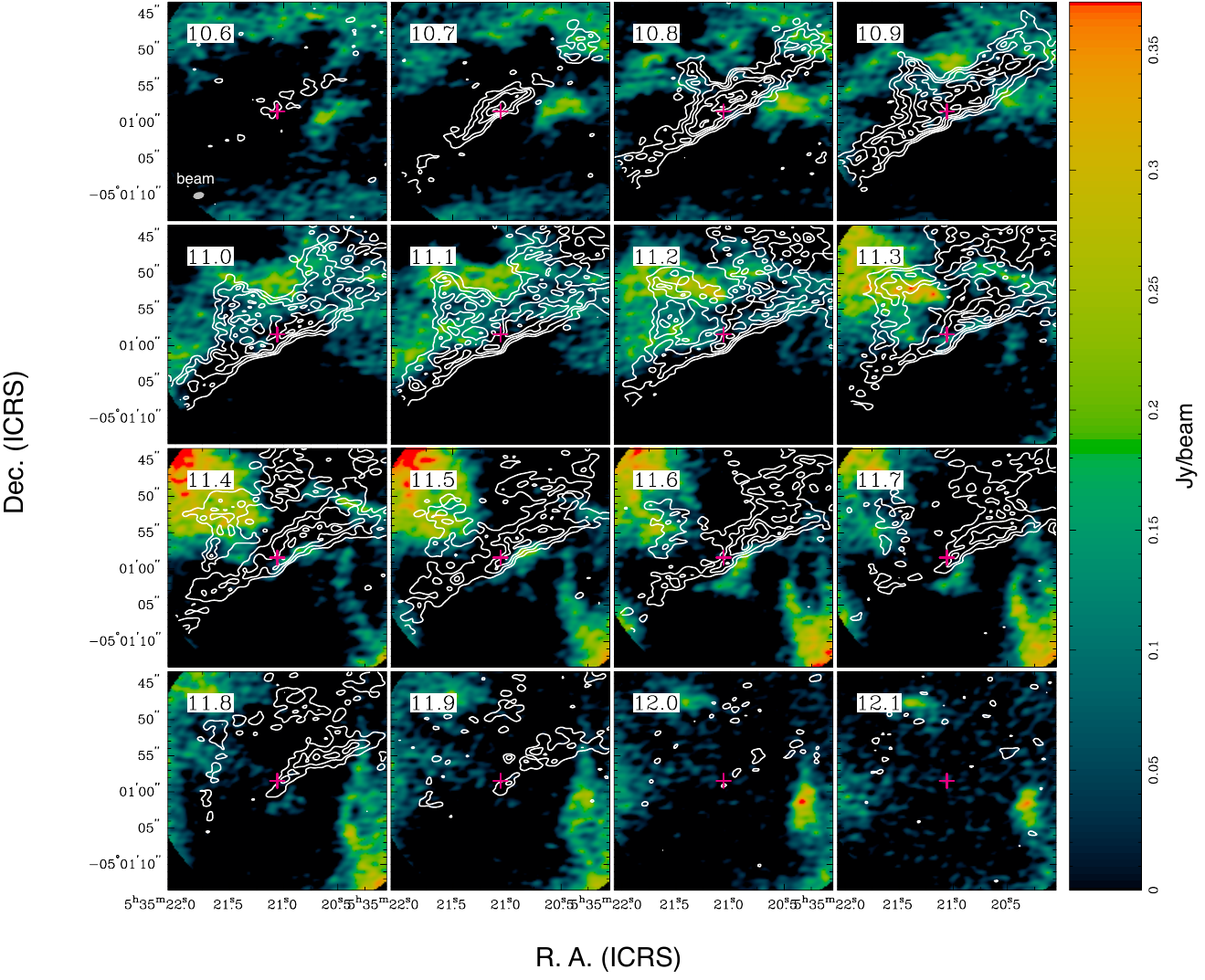}
\caption{Channel maps of the C$^{18}$O 2--1 (color) and N$_2$D$^+$ 3--2 (white contours) emission lines.
The velocity resolution in 0.1 km s$^{-1}$.
The magenta cross denotes the position of the nucleus.
\label{fig:N2Dp_C18O}}
\end{figure*}

Figures \ref{fig:CO_chmap1} and \ref{fig:CO_chmap2} show the CO 2--1 channel maps at 0.32 km s$^{-1}$ resolution.
The channel map of the N$_2$D$^+$ and that of C$^{18}$O overlaid on the N$_2$D$^+$ are shown in Figures \ref{fig:chmap} and \ref{fig:N2Dp_C18O}, respectively.

\newpage

\newpage

\clearpage

\newpage

\newpage

\section{Total power spectra}
\label{sec:app_TP_spectra}
\restartappendixnumbering

\begin{figure}
\epsscale{0.5}
\plotone{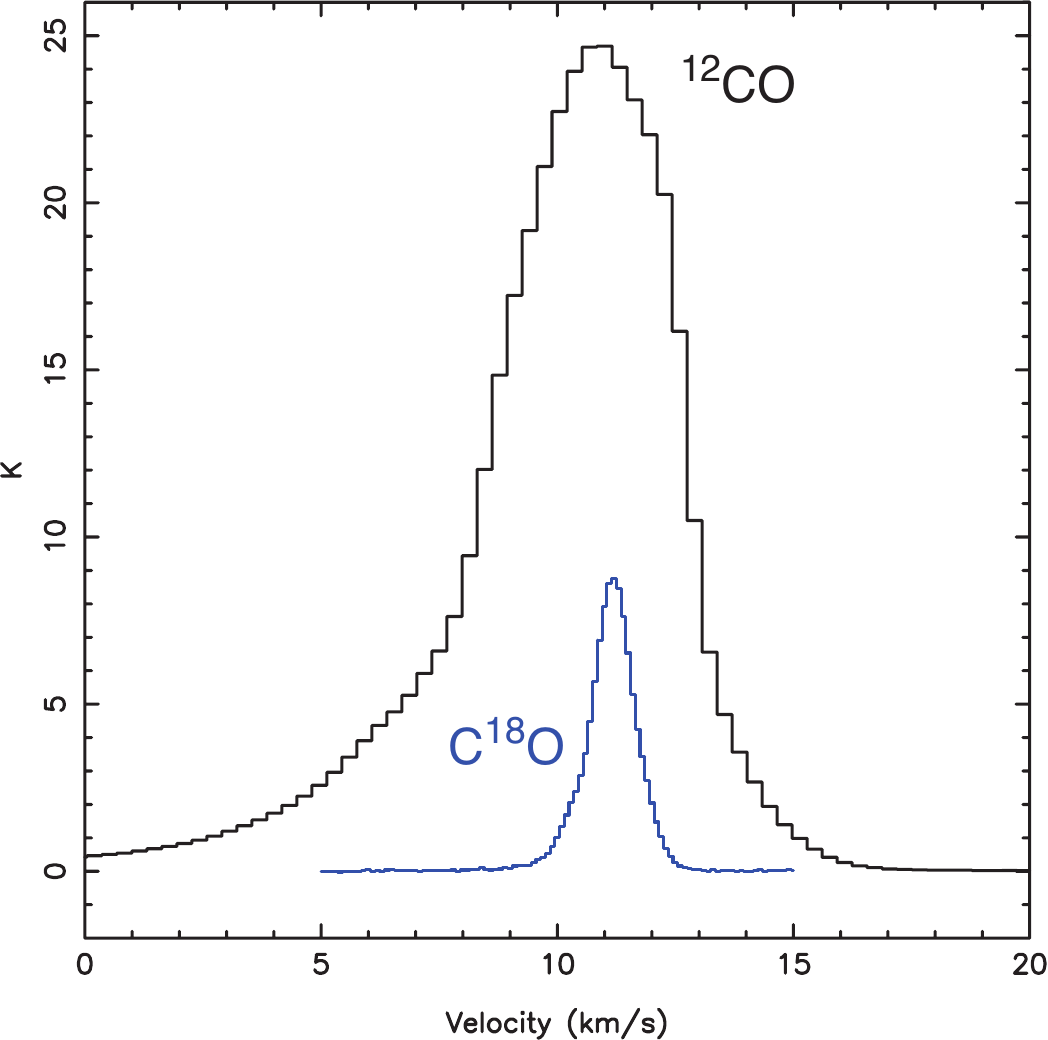}
\caption{$^{12}$CO and C$^{18}$O spectra toward the position of the nucleus obtained with the TP array.
The beam size of the TP array is 28.23\arcsec in $^{12}$CO and is 29.58\arcsec in C$^{18}$O.
\label{fig:TPspectra}}
\end{figure}
\newpage

\bibliography{G208bib}{}
\bibliographystyle{aasjournal}



\end{document}